\documentclass[preprint,12pt]{elsarticle}
\usepackage{amssymb}
\usepackage{graphicx}% Include figure files
\usepackage{dcolumn}% Align table columns on decimal point
\usepackage{bm}% bold math
\usepackage{caption}
\usepackage{float}
\usepackage[dvipsnames]{xcolor}
\journal{Nuclear Physics A}

\begin{document}

\begin{frontmatter}

\title{The nuclear ground-state properties and stellar electron emission rates of $^{76}$Fe, $^{78}$Ni,
$^{80}$Zn, $^{126}$Ru, $^{128}$Pd and $^{130}$Cd using RMF and
pn-QRPA models}
\author{Jameel-Un Nabi$^{1,2}$, Tuncay Bayram$^{3}$, Gul
Daraz$^{2}$, Abdul Kabir$^{2}$ and \c{S}evki \c{S}ent\"{u}rk
$^{3}$}
\address{$^{1}$University of Wah, Quaid Avenue, Wah Cantt 47040, Punjab, Pakistan}
\address{$^{2}$Faculty of Engineering Sciences, GIK Institute of Engineering Sciences and Technology, Topi 23640, Khyber Pakhtunkhwa, Pakistan}
\address{$^{3}$Department of Physics, Karadeniz Technical University, Trabzon, 61080, Turkey}

%%%%%%%%%%%%%%%%%%%%%%%%%%%%%%%%%%%%%%%%%%%%%%%% ABSTRACT  %%%%%%%%%%%%%%%%%%%%%%%%%%%%%%%%%%%%%%%%%%%%%%%%%%%%%%%%%%%%%%%%%%%%%%%%%%%%%%%%%%%%%%%%

\begin{abstract}
Our study consists of investigations of nuclear ground state
properties and weak transition rates of even-even waiting point
nuclei. The calculation was performed for $N=50$ and $N=82$ nuclei.
The Relativistic Mean Field (RMF) model was used to explore the
nuclear ground state properties of selected nuclei. The
proton-neutron quasi particle random phase (pn-QRPA) model was used
for the computation of allowed Gamow Teller (GT) and unique
first-forbidden (U1F) transitions of the selected waiting point
nuclei. The RMF approach with different density-dependent
interactions, DD-ME2 and DD-PC1, was used to compute potential
energy curves and surfaces, quadrupole moments, deformation
parameters, binding energies, proton-neutron separation energies,
charge, and radii. The RMF computed deformation parameters were used
in the pn-QRPA model, as a free parameter, for the
computation of GT and U1F weak transitions. We investigated three
different sets of deformation parameter for the calculation of
electron emission rates. The rates changed considerably with change in
deformation parameter. We later investigated contribution of allowed
GT and U1F rates and competition between positron capture and
electron emission rates at high stellar temperatures. The computed
positron capture rates were significant especially at low densities
and high temperatures. The contribution of U1F rates to allowed GT
appreciably reduced the total calculated half-lives. The comparison
of our results with previous calculations and measurement is also
shown. The pn-QRPA calculation including U1F contribution is in good
agreement with the experimental data.
\end{abstract}

%%%%%%%%%%%%%%%%%%%%%%%%%%%%%%%%%%%%%%%%%%%%%%%%%%%%%%%%%%% End Abstract %%%%%%%%%%%%%%%%%%%%%%%%%%%%%%%%%%%%%%%%%%%%%%%%%%%%%%
\end{frontmatter}
%%%%%%%%%%%%%%%%%%%%%%%%%%%%%%%%%%%%%%%%%%%%%%%%%%%%%%%%%%%%%%%%%%%%%%%%%%%%%%%%%%%%%%%%%%%%%%%%%%%%%%%%%%%%%%%%%%%%%%%%%%%%%%%%%%%%
\section{Introduction}

%%%%%%%%%%%%%%%%%%%%%%%%%%%%%%%%%%%%%%%%%%%%%%%%%%%%%%%%%%%%%%%%%%%%%%%%%%%%%%%%%%%%%%%%%%
The nucleosynthesis mechanism commonly referred to as
rapid neutron-capture process ($r$-process) is believed to synthesize
nearly half of the heavy elements beyond iron \cite{Bur57,Cory61}. Generally, the
$r$-process takes place in a stellar explosive environment with a
large density flux of neutrons ($>10^{20}cm^{-3}$) and high
temperature (T$\approx 10^{9}$K) \cite{Bur57,Cow91,Kra93,Woo94}. It
was observed that all nuclear matter can not be synthesized
simultaneously under the same astrophysical conditions of constant
temperature and density \cite{Kra93}. Under such a scenario, neutron
capture takes place at a much faster rate than competing beta decay.
The $r$-process path proceeds with the larger neutron-flux matter,
that have comparatively small and approximately constant $S_{n}$
(neutron separation energy). At neutron magic shell isotones with
$N$ = 50, 82 and 126, (also termed as waiting points) the
$r$-process flow of matter slows down. The corresponding nuclei have
to wait for several beta decays to occur before capturing of neutrons may resume.
At these magic waiting points, the nuclear matter is accumulated,
resulting in the well-known peaks in the observed $r$-process
abundance distribution.

The study of closed magic shell nuclei is very crucial for a
better understanding of the astrophysical $r$-process. Due to their
stronger binding energies, the neutron separation energies show
discontinuities at these magic shell nuclei
\cite{Arnould07,Arcones11}, resulting in rather low neutron capture
rates. As a consequence the $r$-process matter flow moves closer to
stability, where nuclei have substantially larger EE half-lives.
Thus matter accumulates around closed shell nuclei with N=50, 82,
and 126, producing the observed peaks in the $r$-process distribution. The open shell nuclei become
abundant in comparison to their closed shell counterparts as the shell effects
disappear \cite{Furusawa17}. Our present calculations are only
applicable to even-even closed shell waiting points. However, in
future we will extend our investigations to open shell nuclei in
order to provide a complete coverage for $r$-process simulation
studies.

The half-lives ($T_{1/2}$) of electron emission (EE) calculation
have a considerable impact on the dynamics of $r$-process and
abundance distribution. The calculated $T_{1/2}$ of EE transitions
of waiting points nuclei describes the time scale it takes the mass
flow to transpose seed nuclei to larger ones in the third peak
around $A \sim$ 200. The available data for the EE half-lives of
closed shell waiting points is insufficient \cite{Hos05,Kra86,Pfe01}
and requires attention of theorists and experimentalists.
The radioactive ion beam facilities (e.g.
RIKEN~\cite{Nis11} and FAIR-GSI~\cite{Kur09}) will improve the scenario
in near future. The required half-lives for the simulation of
$r$-process primarily come from theoretical investigations
\cite{Ney20}.

For the investigation of some fundamental nuclear ground state
properties of even-even waiting point isotones with $N$ = 50 and 82,
we employed the Relativistic Mean Field (RMF) approach in axially
deformed shape with different density-dependent interactions. The
RMF model is a phenomenological approach for description of various
ground-state nuclear properties of nuclei and it describes nuclei at
hadronic degrees of freedom. It was first introduced as a fully
fledged quantum field theory by Walecka \cite{walecka1974}. Later,
an additional density dependence has been added in the  RMF theory
for quantitative and correct description of surface properties of
nuclei \cite{boguta1977}. It has been applied to obtain nuclear
properties of nuclei not only in stability but also far from
stability
\cite{ring1996,lalazissis1999,vretenar2005,meng2006,gambhir2006,
bayram2011,bayram2013a,bayram2013b,bayram2018,bayram2019}.

Nuclei with proton number $Z=40-50$ and
neutron number $N=82$ are close to neutron drip-line. We
investigate ground-state nuclear properties for these
isotones by using the RMF model with density-dependent interactions
because RMF model is successful in describing ground-state
properties of nuclei far from stability line. The potential energy
curves (PECs) and the potential energy surfaces (PESs) of selected
nuclei was performed. The quadrupole deformation
parameter, calculated within the framework of RMF model, was
later used in the pn-QRPA model as a free parameter.

We further investigate the allowed Gamow-Teller GT and unique first-forbidden U1F stellar weak electron emission rates ($\lambda_{EE}$)
for $N$ = 50 and 82 isotones by employing the pn-QRPA model with a
deformed Nilsson potential basis. The first attempt to calculate the
microscopic  weak interaction rates for a large number of
available nuclei, far away from the stability line was performed by
Klapdor et al. \cite{Kla84} by using the pn-QRPA approach. Later
on these calculations were refined by Staudt et al.
\cite{Sta89,Sta90} and Hirsh et al. \cite{Hir93} emplying the same nuclear
model. Nabi and Klapdor-Kleingrothaus, for the first time, employed a deformed pn-QRPA model with
schematic separable interaction and calculated the stellar weak rates of
$sd$, $fp$ and $fpg$-shell nuclei under astrophysical conditions of
temperature and density \cite{Nabi99d,Nabi99c,Nabi04}. This approach
provides a state-by-state calculation of weak transition rates
in a microscopic fashion.

For $N$ = 50 and 82 isotones, the past theoretical calculations of
EE rates are overall in decent comparison with available measured
data \cite{Cuenca07,Zhi13}. The important reason for disagreement
between different theoretical calculation is that they did not take
into account the computation of first-forbidden (FF) transitions.
The presence of single-particle energy states with different
parities can lead to a significant impact on calculated half-lives.
The forbidden transitions possess a decent contribution for the $N$
= 50 and 82 isotones. In this regard, Homma et al. \cite{Hom96} made
a first effort to include the contribution of unique first-forbidden
(U1F) weak rates to the total rates by using the pn-QRPA approach.
Borzov, recently computed the FF inclusion to the total weak rates
by using self-consistent density-functional+continuum QRPA approach
\cite{Bor06}. His study shows a substantial reduction of total
half-lives with the inclusion of FF transitions for waiting point
isotones. The FF inclusion was also calculated by using the large
scale shell model (LSSM) \cite{Zhi13}.

The authors in Ref.  \cite{Marketin16} employed fully self-consistent
covariant density functional theory (CDFT) for the ground and
excited states of nuclei. The $\beta$-decay half-lives and neutron
emission probabilities were calculated for a large number of nuclei.
The FF transitions were found to contribute significantly to the total decay
rates.
More recently the FF contribution was studied by using the
deformed pn-QRPA theory for neutron rich nuclide in stellar matter
by \cite{Nabi17}.

In this paper we plan to investigate nuclear structure properties of $N$ = 50 and 82
isotones using the RMF model. We later employ the pn-QRPA model for the computation
of weak rates and half-lives of selected waiting points.

Section~2 describes briefly the necessary formalism. Results of our calculations and
 comparison with previous results and measurement are discussed in Section~3. We finally conclude our findings in Section~4.
%%%%%%%%%%%%%%%%%%%%%%%%%%%%%%%%%% End of Introduction Part %%%%%%%%%%%%%%%%%%%%%%%%%%%%%%%%%%%%%%%%%%%%%%%%%%%%5

\section{Theoretical Framework}

%%%%%%%%%%%%%%%%%%%%%%%%%%%%%%%%%% Description of RMF Model %%%%%%%%%%%%%%%%%%%%%%%%%%%%%%%%%%%%%%%%%%%%%%%%%%%%%5
\subsection{The RMF Model}

{The RMF model based calculations are described in such a way that
nucleons interact with each other with exchange of various mesons
and photon. We only consider $\sigma$, $\omega$ and $\rho$ mesons in
our model. One can find few types of non-linear RMF model based on
handling of self-interaction of mesons and density-dependent
meson-nucleon couplings (for further follow-up see
Ref.~\cite{meng2006} and references therein). The RMF model based
density-dependent meson-nucleon couplings is described shortly in
this subsection. The RMF model starts with phenomenological
relativistic Lagrangian density given as}

\begin{eqnarray}
{\mathcal{L}=\bar\psi(i\gamma\partial-m)\psi+\frac{1}{2}(\partial\sigma)^{2}-\frac{1}{2}m_{\sigma}\sigma^{2}-\frac{1}{4}
{\ \Omega}_{\mu\nu}{\ \Omega}^{\mu\nu}\nonumber}\\
{+\frac{1}{2}m^{2}_{\omega}\omega^{2} -\frac{1}{4}\vec{\
R}^{\mu\nu}\vec{\
R}_{\mu\nu}+\frac{1}{2}m_{\rho}^{2}\vec{\rho}^{2}_{\mu}-
\frac{1}{4}{\ F}^{\mu\nu}{\ F}_{\mu\nu}-g_{\sigma}\bar\psi\sigma\psi \label{lagrangian}}\\
{-g_{\omega}\bar\psi\gamma.\omega\psi-g_{\rho}\bar\psi\gamma.\vec{\rho}\vec{\tau}\psi-e\bar\psi\gamma.A\frac{1-\tau_{3}}{2}\psi.\nonumber}
\label{Lagrangian}
\end{eqnarray}

{The $\psi$ is Dirac spinor and it represents nucleon with mass {\it
m}. The $m_{\sigma}$, $m_{\omega}$ and $m_{\rho}$, are the masses of
$\sigma$, $\omega$ and $\rho$ mesons, respectively. $g_{\sigma}$,
$g_{\omega}$ and $g_{\rho}$ represent the coupling constants for
the mesons to the nucleon. The field tensors of $\omega$, $\rho$ and
the photon were expressed as}

\begin{eqnarray}
{{\bf \Omega}^{\mu\nu}=\partial^{\mu}\omega^{\nu}-\partial^{\nu}\omega^{\mu}, \label{fields:sub1}\nonumber}\\
{\vec{{\bf R}}^{\mu\nu}=\partial^{\mu}\vec{\rho}^{\nu}-\partial^{\nu}\vec{\rho}^{\mu},\label{fields:sub2}}\\
{{\bf F}^{\mu\nu}=\partial^{\mu}A^{\nu}-\partial^{\nu}A^{\mu}.}
\label{fields:sub3}\nonumber
\end{eqnarray}

{In these equations vectors in three-dimensional space are shown with
bold-faced symbols where in the isospin space the vectors are
represented in conventional symbol having overhead arrows. The unknown
meson masses and coupling constants are parameters and they are
fine-tuned by using known experimental data to its correct
predictions of nuclear matter properties and ground-state properties
of finite nuclei. The meson-nucleon vertex functions can be fixed-on
by tuning the parameters of the assumed phenomenological density
dependence of the meson-nucleon couplings to regenerate the
properties of symmetric and asymmetric nuclear matter and finite
nuclei as reported in Ref.~\cite{lalazissis2005}.}

{Tacking into account the classical variational
principle, the equations of motion for the fields can be derived
from the relativistic Lagrangian density given in
Eq.~(\ref{lagrangian}). A set of coupled equations (Dirac equation
with the potential terms for the nucleons) and the Klein-Gordon like
equations with sources for mesons and photon are obtained. The set
of these equations can be solved separately by the expansion of the
wave functions in terms of a spherical, an axially and a triaxially
symmetric harmonic oscillator potential ~\cite{niksic2014}. In the
present study, axially and triaxially symmetric RMF calculations
have been performed to obtain ground-state nuclear properties of
$^{76}$Fe, $^{78}$Ni, $^{80}$Zn, $^{126}$Ru, $^{128}$Pd and
$^{130}$Cd nuclei as well as to obtain their  PECs and PESs by
following the recipe of Ref.~\cite{niksic2014}. In the  present
calculation we employ two different types of density-dependent
interactions namely DD-ME2~\cite{lalazissis2005} and
DD-PC1~\cite{niksic2008}. We took into account 12 shells for
neutrons and protons. No convergence problem was witnessed in our
iterative calculations.}

%%%%%%%%%%%%%%%%%%%%%%%%%%  pn-QRPA Model   %%%%%%%%%%%%%%%%%%%%%%%%%%%%%%%%%%%%%%%%%%%%%%%%%%%%%%%%%%%%%%%%%%%%%%%%%%%%
\subsection{The pn-QRPA model}

Our calculation involves solution of pn-QRPA equations in a
multi-shell single-particle space with a schematic interaction. The
Hamiltonian of the model is given by
\begin{equation}
H^{QRPA} = H^{sp} + V^{pair} + V ^{ph}_{GT} + V^{pp}_{GT}.
\label{Eqt. Hamiltonian}
\end{equation}

Single particle energies and wave functions were calculated in the
deformed Nilsson potential basis. Pairing was treated within the BCS
formalism. The calculated EE transitions and related half-lives
strongly depend on the Q-values and residual interactions
\cite{Engel99}. The residual interactions $\chi$ ($\textit{ph}$) and
$\kappa$ ($\textit{pp}$), known as particle-hole part and
particle-particle part respectively, were considered for the
computation of both allowed GT and U1F calculation. For a detailed description of $\chi$ and $\kappa$ and optimum choice of these parameters we refer to
 \cite{Sta90,Hir93,Hom96}. For the U1F
transitions, the $pp$ and $ph$ matrix elements are given by

\textbf{\begin{equation} \label{vph}
V^{ph}_{pn,p^{\prime}n^{\prime}} = +2\chi_{U1F}
f_{pn}(\mu)f_{p^{\prime}n^{\prime}}(\mu),
\end{equation}}

\textbf{\begin{equation}\label{vpp} V^{pp}_{pn,p^{\prime}n^{\prime}}
= -2\kappa_{U1F} f_{pn}(\mu)f_{p^{\prime}n^{\prime}}(\mu),
\end{equation}}

where \textbf{\begin{equation}\label{fpn}
f_{pn}(\mu)=<p|t_{-}r[\sigma Y_{1}]_{2\mu}|n>,
\end{equation}}
represents the single particle transition amplitude between
Nilsson single particle states. $\mu$ represents the spherical
component of the transition operator and takes the
value of $0,\pm1$ and $\pm2$.  The neutron and proton states possess different parities. Other input parameters for computation of weak
transitions are pairing gaps ($\Delta_{p}$, $\Delta_{n}$), nuclear
deformation ($\beta_{2}$) values, Q-values and Nilsson potential parameters
(NPP).  We performed our calculation by using the nuclear
deformation parameter from the recent global calculation of finite
range droplet model (FRDM) \cite{Mol2012}. The NPP were adopted
from \cite{Rag84} and the oscillation constant (identical for both
of protons and neutrons) was determined using the relation
$\hbar\omega=41A^{-{1}{3}}$ [MeV]. We used the Nilsson
potential for the calculation of wave function. The Nilsson model
has been widely used to describe the structure of low-lying states. The nuclear deformation parameter $\beta_{2}$ was
used as an input parameter in the Nilsson potential in our computation. As a
starting point, single particle energies and wave functions were
computed in the deformed Nilsson basis. The transformation from the spherical nucleon
basis (\textit{c}$_{jm}^{+}$,\textit{ c}$_{jm}$) to the axial
symmetric deformed basis
(\textit{d}$_{m\alpha}^{+}$,\textit{d}$_{m\alpha}$) were performed using
\begin{equation}\label{ND}
d_{m\alpha}^{+}=\sum_{j}D_{j}^{m\alpha}c_{jm}^{+},
\end{equation}
where \textit{d}$^{+}$ and \textit{c}$^{+}$ indicate the
particle creation operators in the deformed and spherical
basis, respectively.  The transformation matrix
\textit{D}$_{j}^{m\alpha}$, is a set of Nilsson eigen-functions with
$\alpha$ as an additional quantum number which describes the Nilsson
eigen-states. The BCS formalism was used in the Nilsson basis for
neutron/proton system separately. The transformation matrices were
determined by diagonalization of the Nilsson Hamiltonian (for complete
description we refer to \cite{Hir93, Mut89}).

Pairing gap values, $\Delta_{n}$=$\Delta_{p}$=12$/\sqrt{A}$ [MeV],
was used in our calculation according to the global systematic. The
Q-values were taken from the recent atomic mass data evaluation of
\cite{Audi2017}.

Details of solution of the Hamiltonian shown in Eq.~(\ref{Eqt.
Hamiltonian}) may be studied from \cite{Mut92}. Calculation of
terrestrial EE half-lives may be seen from \cite{Sta90}. Also, the
formalism used for the estimation of allowed GT and U1F transitions
in stellar scenario by using pn-QRPA approach can be studied in more
details from Refs. \cite{Nabi99c,Nabi04,Nabi16,Nabi16B}.

The allowed positron capture (PC) rates from the parent nucleus
$\mathit{i}$th-state to the daughter nucleus $\mathit{j}$th-state is
specified by

\begin{equation}\label{pc}
\lambda _{ij}^{(PC)} =\frac{\ln2}{D} [f_{ij}(T,\rho,E_{f})]
 [B(F)_{ij}+(\frac{g_{A}}{g_{V}})^{2} B(GT)_{ij}].
\end{equation}

The value of $D$ is taken as $6143$~s \cite{Tow09}. In above
equation $(B_{ij})s$ depict the reduced transition probabilities of
Fermi ($B(F)$) and GT ($B(GT)$) interactions. We took the
value of $g_{A}/g_{V}$ (ratio of weak axial and vector coupling
constant)  to be $-1.2694$ \cite{Nak10}. The same value was taken for the calculation of weak transition rates of neutron rich Cu-isotopes in stellar matter. The authors in Ref. \cite{Zhi13} used the value $g_{A}/g_{V}$ = -1.2701 for the closed magic shell nuclei of N= 50, 82 and 126. Detail formalism for computation of allowed capture rates in stellar environment can be studied in \cite{Nabi99c}.

The U1F stellar EE and PC rates from $\mathit{i}$th-state of the
parent to $\mathit{j}$th-state of the daughter nucleus is specified
by

\begin{equation}\label{EE}
\lambda_{ij}^{EE(PC)} =
\frac{m_{e}^{5}c^{4}}{2\pi^{3}\hbar^{7}}\sum_{\Delta
J^{\pi}}g^{2}f_{ij}(\Delta J^{\pi})B_{ij}(\Delta J^{\pi}),
\end{equation}

where $f_{ij}(\Delta J^{\pi})$ is the integrated Fermi function and
$B_{ij}(\Delta J^{\pi})$ is the reduced transition probability for
weak EE and capture rates. ($g$) in the expression is the weak
coupling constant. The weak coupling constant ($g$) take the values
of $g_{V}$ or $g_{A}$ based on $\Delta J^{\pi}$ transitions. The
dynamics part of the rate equation is given by

\begin{equation}\label{e}
B_{ij}(\Delta
J^{\pi})=\frac{1}{12}\zeta^{2}(w_{m}^{2}-1)-\frac{1}{6}\zeta^{2}w_{m}w+\frac{1}{6}\zeta^{2}w^{2},
\end{equation}
where $\zeta$ is
\begin{equation}\label{efg}
\zeta=2g_{A}\frac{\langle
f|\vert\sum_{k}r_{k}[\textbf{C}^{k}_{1}\times
{\sigma}]^{2}{\textbf{t}}^{k}_{-}\vert|i\rangle}{\sqrt{2J_{i}+1}},
\end{equation}
and
\begin{equation}\label{gkl}
\textbf{C}_{lm}=\sqrt{\frac{4\pi}{2l+1}}\textbf{Y}_{lm},
\end{equation}

with $\textbf{Y}_{lm}$ the spherical harmonics. The kinematics
portion of Eq.~(\ref{EE}) can be obtained as
\begin{equation}\label{f}
f_{ij} = \int_{1}^{w_{m}} w \sqrt{w^{2}-1}
(w_{m}-w)^{2}[(w_{m}-w)^{2}F_{1}(Z,w) \nonumber\\
+ (w^{2}-1)F_{2}(Z,w)] (1-D_{-}) dw,
\end{equation}

the term ($w$) relates to the overall kinetic energy of the electron
with the inclusion of its rest-mass energy. The mathematical
expression $ w_{m} = m_{p}-m_{d}+E_{i}-E_{j}$ represents the total
energy of corresponding weak electron emission-decay. Here the terms
$m_{p}$ and $E_{i}$, relates the mass and excitation energies
($E_{xp}$) of the parent nucleus, respectively. Whereas, $m_{d}$ and
$E_{j}$ are the related mass and excitation energies of the daughter
nucleus, respectively. The distribution function for the electrons
is shown by $D_{-}$. We assume that electrons are not residing in a
bound state, and following the Fermi-Dirac distribution functions,

\begin{equation}\label{Gm}
D_{-} = [exp (\frac{E-E_{f}}{kT})+1]^{-1}.
\end{equation}

In the above expression, $E=(w-1)$ and $E_{f}$ indicates the K.E
(kinetic energy) and fermi energy ($E_{f}$) of electrons,
respectively. $T$ is the temperature in units of Kelvin-scale and
the term $k$ implies for the Boltzmann constant. We refer to and
study the recipe of \cite{Gov71} for calculation of Fermi functions,
$F_{1}(\pm Z,w)$ and $F_{2}(\pm Z,w)$ shown in equation
(Eq.~(\ref{f}).

Due to the presence of high temperatures in stellar core, EE and
capture rates have a small contribution from parent excited energy
levels. We use the Boltzmann distribution function to calculate the
occupation probability of parent $ith$-state is

\begin{equation}\label{pi}
P_{i} = \frac {exp(-E_{i}/kT)}{\sum_{i=1}exp(-E_{i}/kT)}.
\end{equation}

The total stellar EE and PC rates were finally calculated by using
\begin{equation}\label{lb}
\lambda^{EE(PC)} = \sum_{ij}P_{i} \lambda_{ij}^{EE(PC)}.
\end{equation}

The summation stands for computation of all parent and daughter
energy levels until required and desire convergence is achieved. We
found that due to the presence of large model space (about up to
$7\hbar\omega$ major oscillatory shells) desire convergence can be
easily obtained in our weak EE and PC computations. This is one of
the big success of pn-QRPA approach as it may be employed for
computation of weak rates of any arbitrary heavy nuclear-species.

%%%%%%%%%%%%%%%%%%%%%%%%%%%%%%%%%%%%%%%%%%%%%%%%%%%%%%%%%%%%%%%%%%%%%%%%%%%%%%%%%%%%%%%%%%%%%%%%%%%%%%%%%%%%

\section{Results and discussion}

%%%%%%%%%%%%%%%%%%%%%%%%%%%%%%%% Results and discussion Part of RMF-Model   %%%%%%%%%%%%%%%%%%%%%%%%%%%%%%%%%%

\begin{figure}[h]
\centering
\includegraphics[width=0.8\textwidth]{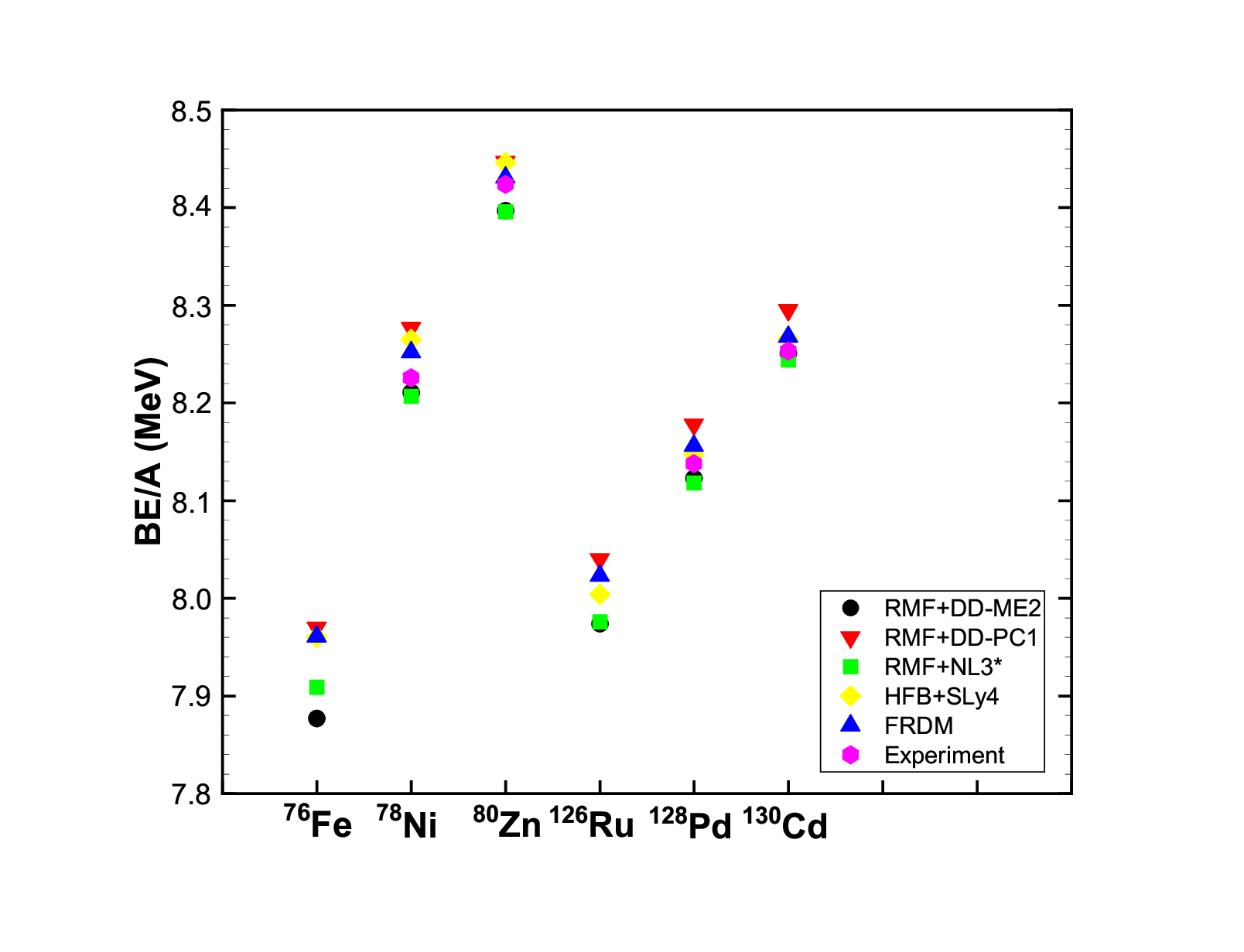}
\caption{Calculated binding energies per nucleon for $^{76}$Fe,
$^{78}$Ni, $^{80}$Zn, $^{126}$Ru, $^{128}$Pd and $^{130}$Cd using the RMF model with DD-ME2 and DD-PC1 interactions. Results of the RMF model with NL3* interaction~\cite{bayram2013a}, HFB
theory with Sly4 parameter set~\cite{stoitsov2003},
FRDM~\cite{Mol2012} and available experimental
data~\cite{wang2012} are also shown for comparison.} \label{bea}
\end{figure}

{We first discuss the results
obtained from the axially deformed RMF-model with different density
interaction parameters. The calculated binding energies of $N$ = 50
 ($^{76}$Fe, $^{78}$Ni, $^{80}$Zn) and $N$ = 82 ($^{126}$Ru, $^{128}$Pd
$^{130}$Cd) nuclei by using RMF model with DD-ME2 and DD-PC1
interactions are shown in Fig.~\ref{bea}. The results of binding
energies for these nuclei employing the RMF model with non-linear NL3*
interaction parameters \cite{bayram2013a}, Hartree-Fock Bogoliubov
(HFB) theory with Sly4 parameter set \cite{stoitsov2003},
finite-range droplet model (FRDM) \cite{Mol2012} and available
experimental data \cite{wang2012} are also shown for comparison.
Maximum deviations of binding energies from available experimental
data ($^{78}$Ni, $^{80}$Zn, $^{128}$Pd and $^{130}$Cd nuclei) for
RMF[DD-ME2], RMF[DD-PC1], RMF[NL3*], HFB[Sly4] and FRDM are between
$0.024-0.049$ MeV, $0.004-0.027$ MeV, $0.011-0.027$ MeV,
$0.006-0.037$ MeV and $0.012-0.025$ MeV, respectively. The
results indicate that the ground-state binding energies of the
nuclei obtained from the RMF model with DD-ME2 interaction are much
closer to the experimental results.}

\begin{figure}[h]
\centering
\includegraphics[width=0.55\textwidth]{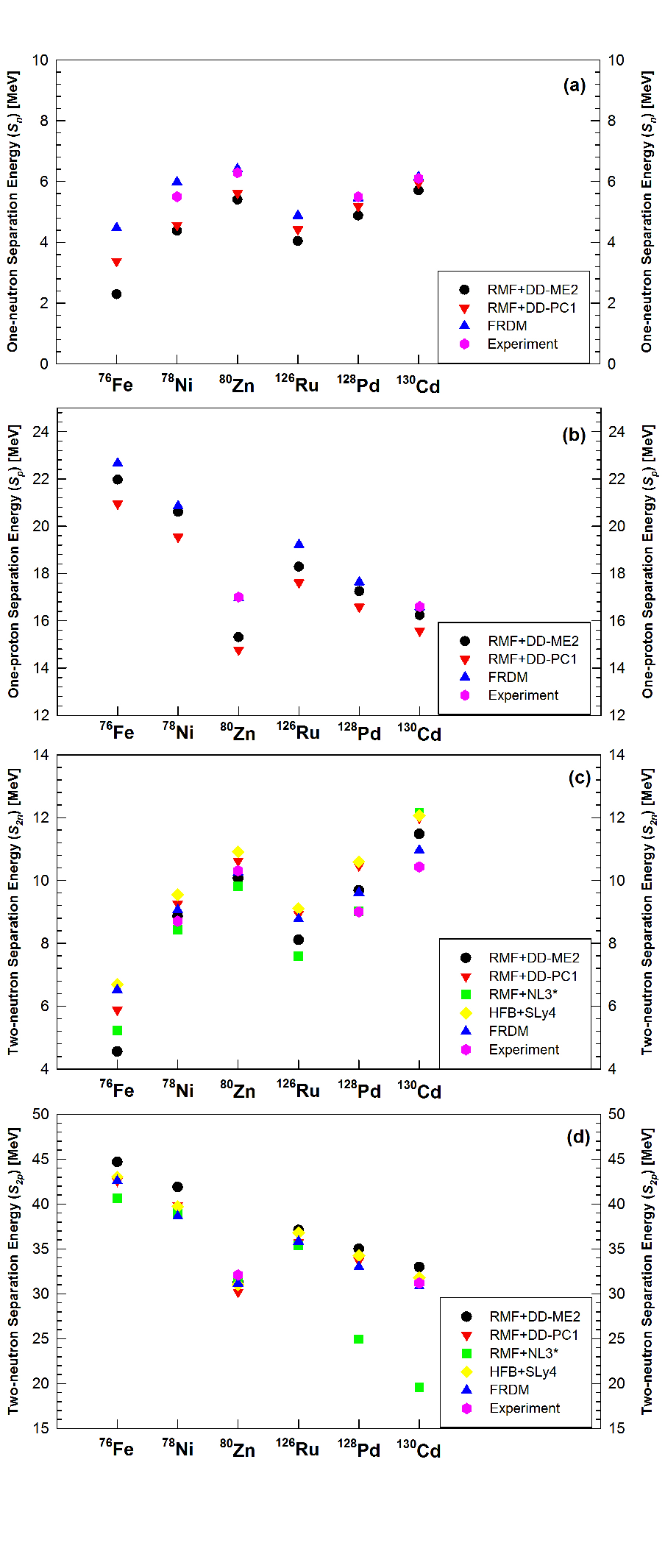}
\caption{Calculated one-neutron (a), one-proton (b), two-neutron
(c) and two-proton (d) separation energies for $^{76}$Fe, $^{78}$Ni,
$^{80}$Zn, $^{126}$Ru, $^{128}$Pd and $^{130}$Cd using the RMF
model with DD-ME2 and DD-PC1 interactions. Available
theoretical predictions from the RMF model with NL3*
interaction~\cite{bayram2013a}, HFB theory with Sly4 parameter
set~\cite{stoitsov2003}, FRDM~\cite{Mol2012} and experimental
data~\cite{wang2012} are also shown for comparison.} \label{Sall}
\end{figure}

{Nucleon separation energies for one-neutron ($S_{n}$), one-proton
($S_{p}$), two-neutron ($S_{2n}$) and two-proton ($S_{2p}$) for the
corresponding waiting points by using RMF model with DD-ME2 and
DD-PC1 interactions are shown in Fig.~\ref{Sall}. The results of RMF
model with NL3* interaction \cite{bayram2013a}, HFB theory with SLy4
interaction \cite{stoitsov2003}, FRDM \cite{Mol2012} and
experimental data \cite{wang2012} are also shown in Fig.~\ref{Sall}
for comparison. As seen in Fig.~\ref{Sall}a and Fig.~\ref{Sall}b,
the results of FRDM for $S_{n}$ and $S_{p}$ separation energies of
selected nuclei are much closer to available experimental data. The
maximum deviation between the experimental data and the results of
RMF model with DD-ME2 and DD-PC1 interactions is under 1 MeV. The
same success of FRDM model in describing $S_{2n}$ and $S_{2p}$
energies of selected nuclei is seen in Fig.~\ref{Sall}c and
Fig.~\ref{Sall}d. On the other hand, the RMF model with
DD-ME2, DD-PC1 and NL3* interactions and HFB theory with SLy4
interaction provide compatible results.

%On the other hand, the RMF model with DD-ME2 and DD-PC1 interactions
%and HFB theory with SLy4 interaction provide compatible results.
%However, the RMF model with NL3* interaction gives slightly
%different results among the others.
}

\begin{figure}[h]
\centering
\includegraphics[width=1.0\textwidth]{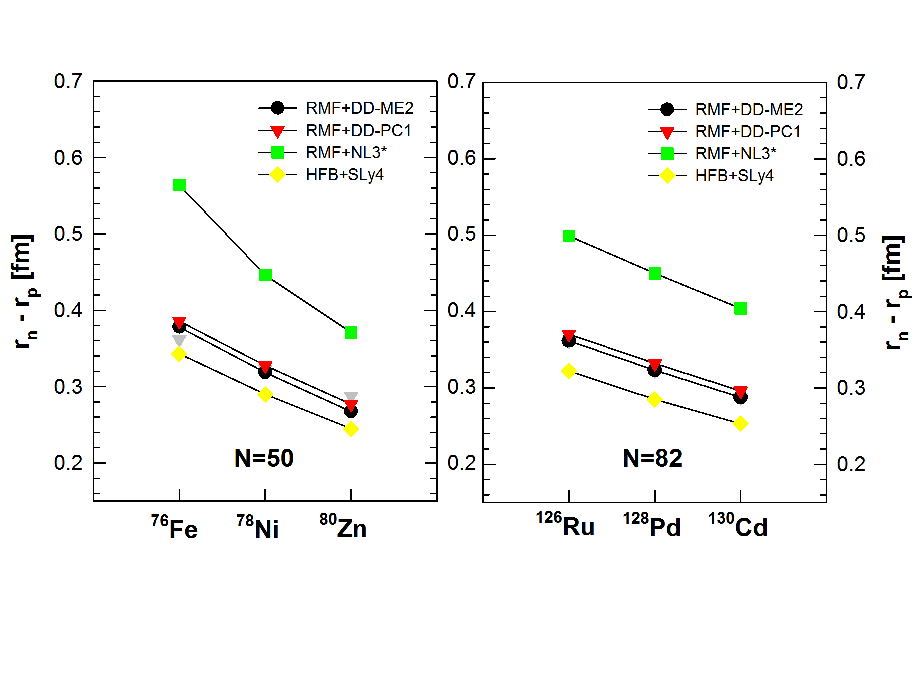}
\caption{Calculated neutron skin thickness for $^{76}$Fe,
$^{78}$Ni, $^{80}$Zn, $^{126}$Ru, $^{128}$Pd and $^{130}$Cd
using the RMF
model with DD-ME2 and DD-PC1 interactions.  Results of the RMF model with NL3*
interaction~\cite{bayram2013a} and HFB theory with Sly4
interaction~\cite{stoitsov2003} are also shown for comparison.}
\label{rnrp}
\end{figure}

{Accurate knowledge of neutron skin thickness
($\Delta r_{np}=r_{n}-r_{p}$) is important for problems in the field of nuclear physics and
astrophysics. As an example case, the slope parameter of symmetry
energy of nuclear matter is correlated with the neutron skin
thickness \cite{dong2015}. In Fig.~\ref{rnrp}, calculated neutron
skin thickness of $^{76}$Fe, $^{78}$Ni, $^{80}$Zn, $^{126}$Ru,
$^{128}$Pd and $^{130}$Cd nuclei by using RMF model with DD-ME2 and
DD-PC1 interactions are shown. The predictions of RMF model
with NL3* parameter set \cite{bayram2013a} and HFB theory with SLy4
interaction \cite{stoitsov2003} are shown for comparison. As it is
well known from our fundamental nuclear physics knowledge, neutron
skin thickness decreases by adding proton in an isotonic chain of
nuclei. This case is clearly visible for $N=50$ and $N=82$ isotonic
chains in Fig.~\ref{rnrp}. It may be understood from the
figure that the RMF model and HFB theory give close results to each other
for neutron skin thickness of the selected nuclei. It may be further
noted that the results of RMF model with NL3* differ
from those of others while its tendency for adding protons is
same.}

{Experimentally, the root-mean-square (rms) charge radius of
a nucleus is measured by using the interactions between the nucleus
and electrons or muons. One can find experimental methods
for determination of rms charge radii of a nucleus along stability line
and far away from the stability line \cite{fricke1995}. Latest
compiled experimental charge radii data of nuclei can be found in
Ref. \cite{angeli2013}. There is no available
experimental data for rms charge radii of $^{76}$Fe, $^{78}$Ni,
$^{80}$Zn, $^{126}$Ru, $^{128}$Pd and $^{130}$Cd because these nuclei
are close to the neutron dripline. On the other hand some empirical
charge radii formulae, where parameters are fitted by using
known experimental data, works well \cite{bayram2013c}. One of the
latest refitted empirical charge radii formulas by using the data of
898 nuclei is}
\begin{equation}
{\label{rc} R_{c}=\sqrt{5/3}((r_{p}Z^{1/3})^{2} + 0.64)^{1/2}},
\end{equation}
{where $r_{p}$ and $Z$ are proton radii and proton numbers of
nuclei, respectively. This formula gives the smallest
root-mean-square deviation with respect to other empirical charge
radii formulae. Details can be found in Ref. \cite{bayram2013c}. In
Table~\ref{charge}, the calculated nuclear charge radii of $N=50$
and $N=82$ nuclei by using the RMF model with DD-ME2 and DD-PC1
interactions are listed. The results of RMF[NL3*], HFB with the
Skyrme force SLy4 and empirical formula are further listed for
comparison. Calculated charge radii of selected nuclei within the
framework of relativistic and non-relativistic mean field methods
are close to each other. It should be noted that the calculated
charge radii values of selected nuclei using the RMF[DD-ME2]
interaction are  closer to the results of latest
refitted empirical charge radii formula given in Eq.~(\ref{rc}).}

{In the present study one may conclude that the RMF
model with density-dependent DD-ME2 and DD-PC1 interactions gives
close results to the available experimental data for the
ground-state nuclear properties of the selected nuclei. The
calculated ground-state nuclear properties of $^{76}$Fe, $^{78}$Ni,
$^{80}$Zn, $^{126}$Ru, $^{128}$Pd and $^{130}$Cd using the RMF
model, with DD-ME2 and DD-PC1 interactions, are listed in
Table~\ref{ddme2} and Table~\ref{ddpc1}, respectively. The
calculated binding energies, radii of neutron, proton, and charge,
deformation parameters and total quadrupole moments ($Q_{T}$) for
selected nuclei are listed in these tables. The calculated
($S_{n}$, $S_{p}$, $S_{2n}$ and $S_{2p}$)-separation energies of
corresponding waiting points are also listed. The results indicate that
these nuclei are close to neutron dripline. It should be noted that
the minus sign in the column of $Q_{T}$ indicates that this nuclei
has oblate shape in its ground-state.}

\begin{figure}[h!t]
\centering
\includegraphics[width=1.0\textwidth]{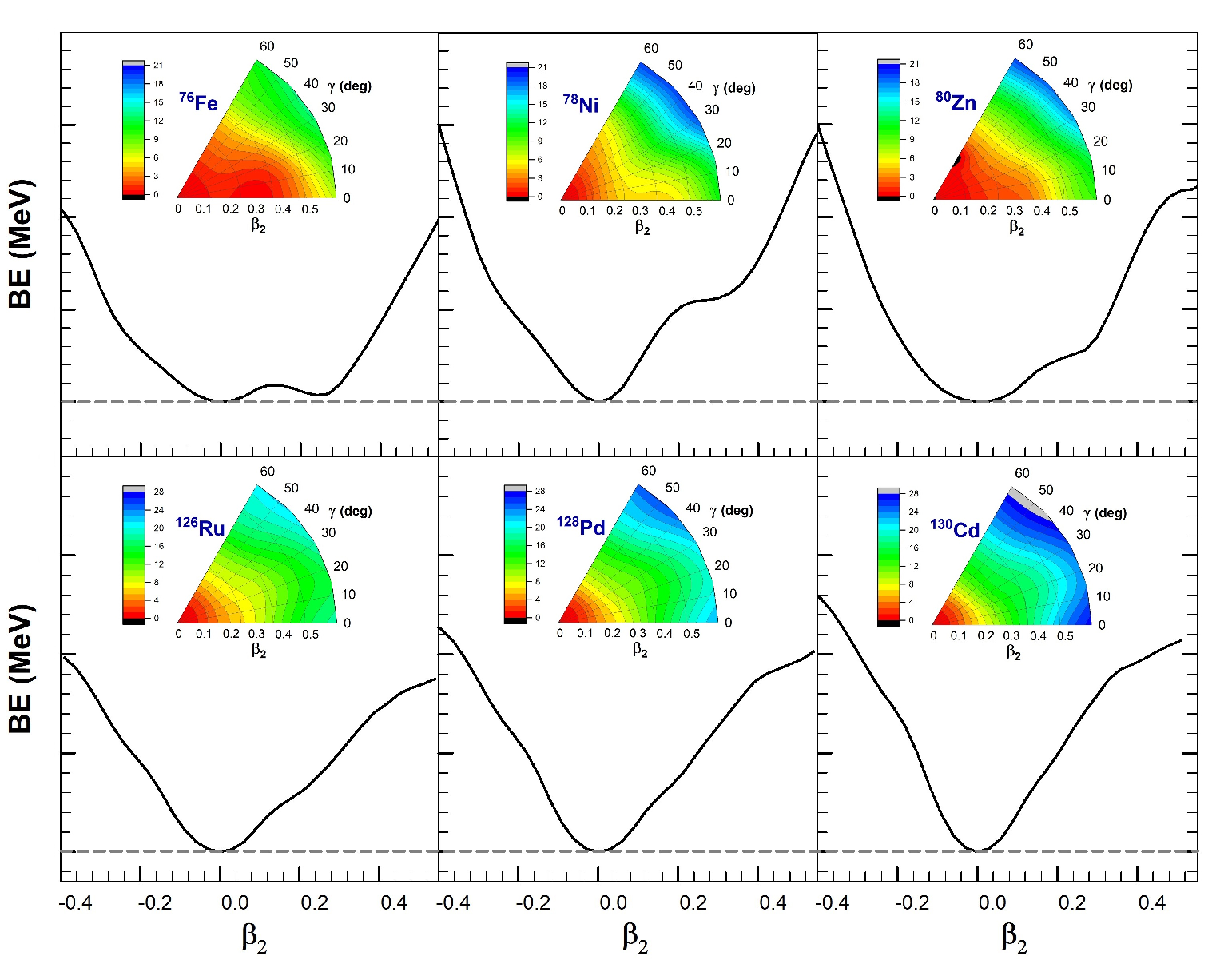}
\caption{Calculated PECs and PESs for $^{76}$Fe,
$^{78}$Ni, $^{80}$Zn, $^{126}$Ru, $^{128}$Pd and $^{130}$Cd using the RMF
model with DD-ME2 interaction.}
\label{pec_pes_me2}
\end{figure}

\begin{figure}[h!t]
\centering
\includegraphics[width=1.0\textwidth]{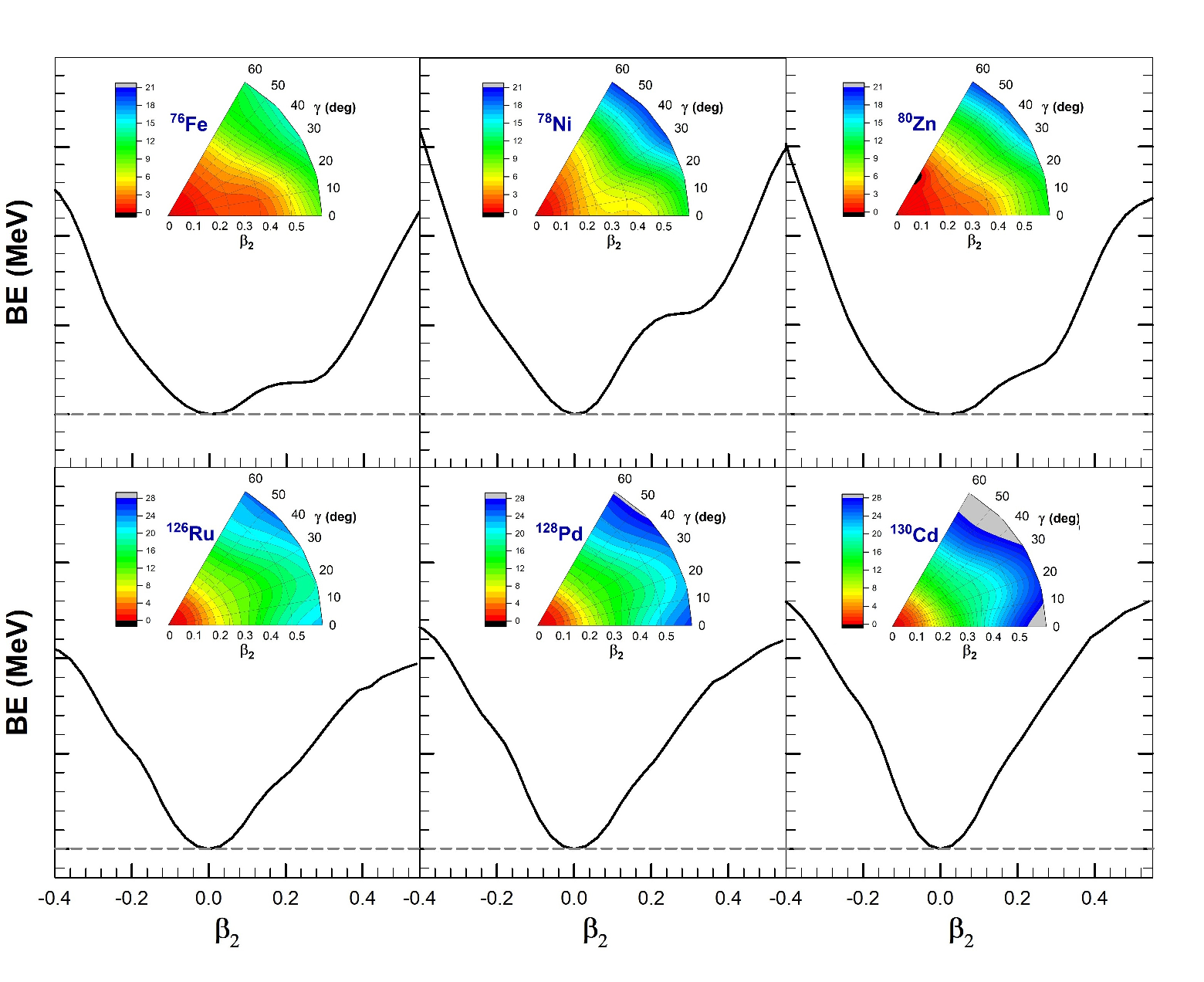}
\caption{Calculated PECs and PESs for $^{76}$Fe, $^{78}$Ni,
$^{80}$Zn, $^{126}$Ru, $^{128}$Pd and $^{130}$Cd  using the RMF
model with DD-PC1 interaction.} \label{pec_pes_pc1}
\end{figure}

{In the present study, we have used the  calculated nuclear deformation
obtained from the RMF model in our pn-QRPA model as input parameter for the selected
nuclei. Because of this reason, we studied the ground-state
deformation parameters of these nuclei in detail. For this
purpose we have carried out PECs and PESs of the selected nuclei
in the RMF model. For PEC calculation, we imposed constraint on the
quadrupole moment to calculate binding energy. Later, lowest
binding energy was used as reference. Thus, differences between
calculated binding energy for specific $\beta_2$ value and reference
binding energy were obtained for carrying out PEC of the related
nuclei as a function of $\beta_2$.  For obtaining PESs of the
selected nuclei triaxial shape, both $\beta$ and
$\gamma$ deformations were taken into account. In this case
constraints were used for both of the deformations and the
potential energy surfaces in the {$\beta – \gamma$} plane ($0^o
\leq \gamma \leq 60^o$). The PECs and PESs of $^{76}$Fe, $^{78}$Ni,
$^{80}$Zn, $^{126}$Ru, $^{128}$Pd and $^{130}$Cd,  obtained by
using DD-ME2 and DD-PC1 interactions in the RMF model, are shown in
Fig.~\ref{pec_pes_me2} and  Fig.~\ref{pec_pes_pc1}, respectively.
Contour lines represent a step of 0.75 MeV and the binding energy is
set to zero at the minimum of each surface. In our calculations
based on the DD-ME2 and DD-PC1 functionals, the PECs of the selected nuclei show a minima around $\beta_2=0$ and the same
results are seen in the PESs of the nuclei. These results are
consistent with the values of $\beta_2$ given in Table~\ref{ddme2}
and Table~\ref{ddpc1}. From this result it may be concluded that
RMF model predicts the shapes of $^{76}$Fe, $^{78}$Ni, $^{80}$Zn,
$^{126}$Ru, $^{128}$Pd and $^{130}$Cd to be almost
spherical.}

%%%%%%%%%%%%%%%%%%%%%%%%%%%%%%%%%%%%%%%%%%%%%%%%%  Results and Discussions Part Of pn-QRPA Theory  %%%%%%%%%%%%%%%%%%%%%%%%%%%%%%%%%%%%%%%%%%%%%%%%%%%%%%%
After analysis of nuclear ground state properties we next discuss
the calculation of weak rates obtained by employing the pn-QRPA
model. EE weak rates and EE half-lives of $^{76}$Fe, $^{78}$Ni, $^{80}$Zn,
$^{126}$Ru, $^{128}$Pd and $^{130}$Cd for a wide range of stellar
temperature and density were calculated including allowed GT and U1F
transitions. The predictive power of our model becomes more effective for smaller half-life values~\cite{Hir93,Nabi16} and
justifies the implementation of present pn-QRPA model for EE
calculation.

\begin{figure}[h!t]
\centering
\includegraphics[width=1.0\textwidth]{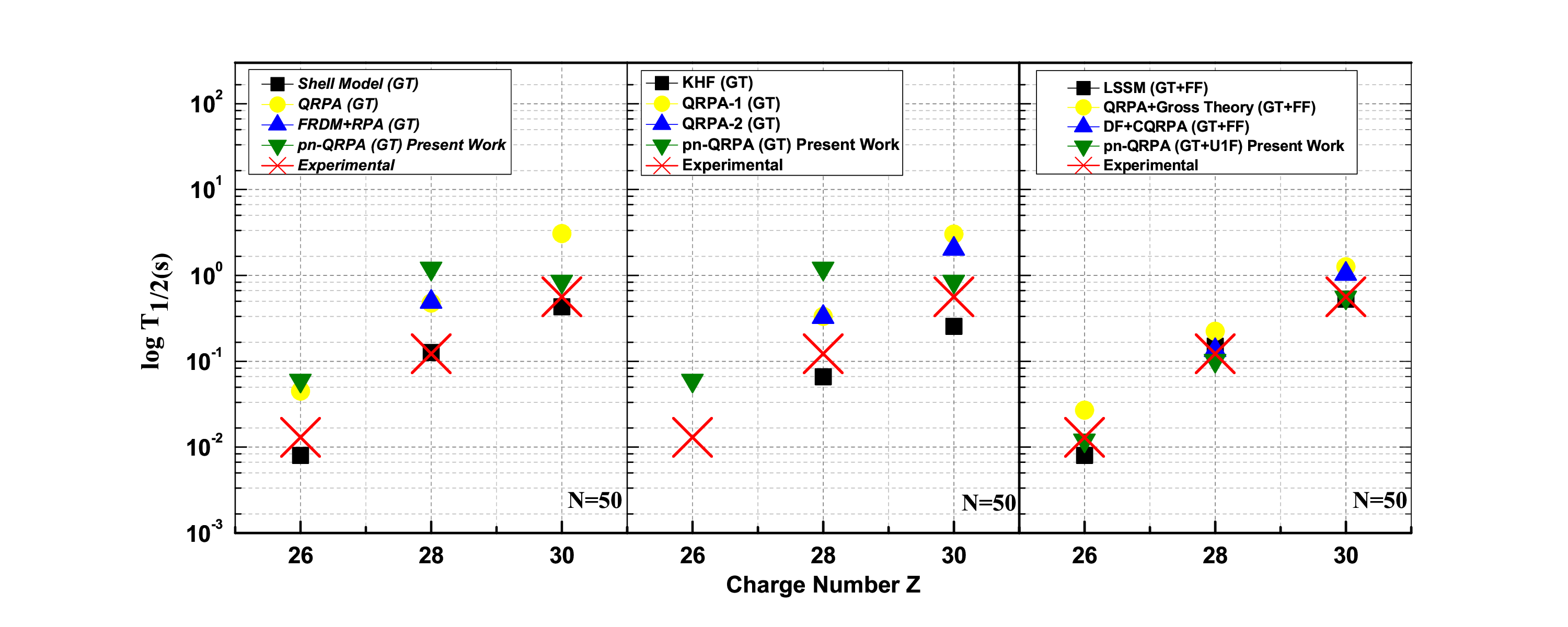}
\caption{Calculated and measured electron emission half-lives of
$N$= 50 waiting point isotones. The right panel depicts the results
of Shell Model (GT) \cite{Lank03}, QRPA (GT) \cite{Mol03} and
FRDM+RPA (GT) \cite{Moller1997}, the middle panel shows the
theoretical computations of KHF, QRPA-1 and QRPA-2 \cite{Pfe02}. The
right panel depicts the results of LSSM (GT+FF) \cite{Zhi13},
QRPA+Gross theory (GT+FF) \cite{Mol03} and DF-CQRPA (GT+FF)
\cite{Bor05}. Allowed GT pn-QRPA calculation is shown as GT while
those including unique first-forbidden contribution as (GT+U1F).
Experimental values were taken from Ref. \cite{Audi2017}.}
\label{50}
\end{figure}

\begin{figure}[h!t]
\centering
\includegraphics[width=1.0\textwidth]{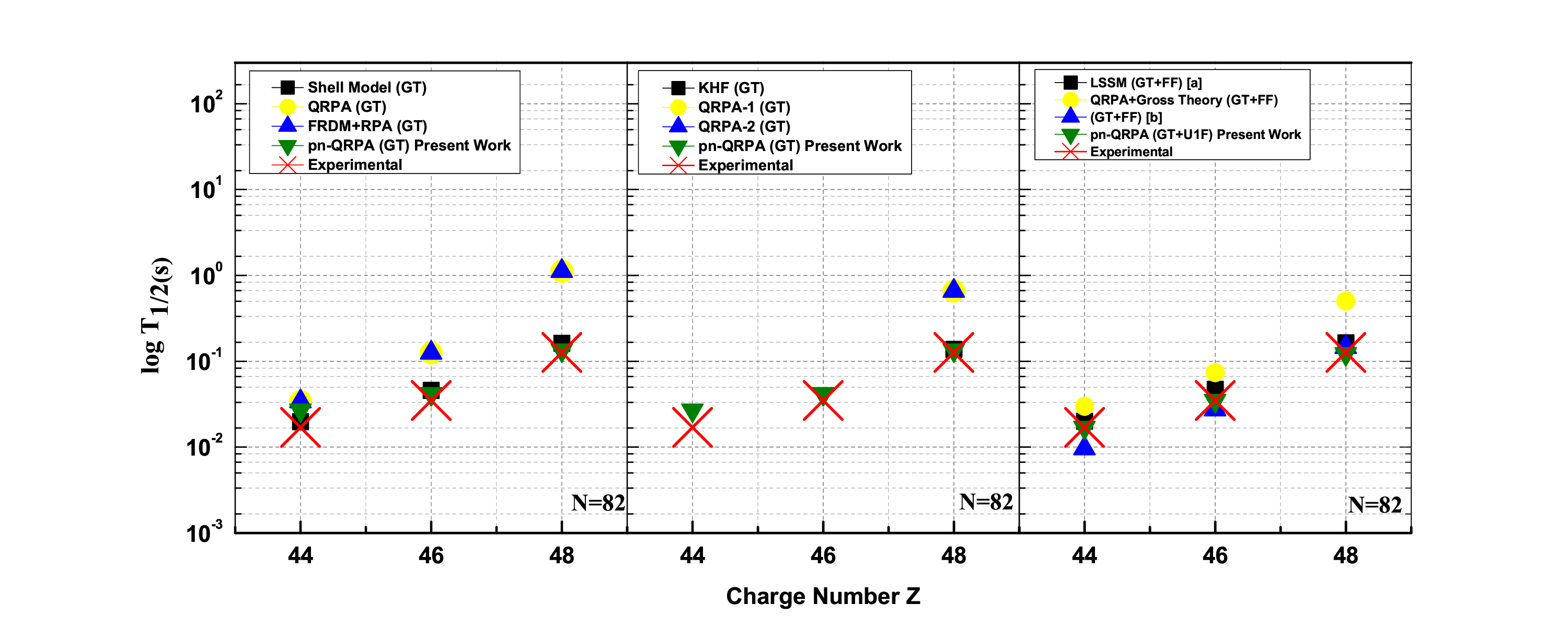}
\caption{Calculated and measured electron emission half-lives of
$N$= 82 waiting point isotones. The right panel depicts the results
of Shell model (GT) \cite{Cuenca07}, QRPA (GT) \cite{Mol03} and
FRDM+RPA (GT) \cite{Moller1997}, the middle panel shows the
theoretical calculations of KHF, QRPA-1 and QRPA-2 \cite{Pfe02}. The
right panel shows the results of LSSM (GT+FF) [a]$\rightarrow$
\cite{Zhi13}, QRPA+Gross theory (GT+FF) \cite{Mol03} and GT+FF
results [b]$\rightarrow$ \cite{Mar99}. Allowed GT pn-QRPA
calculation is shown as GT while those including unique
first-forbidden contribution as (GT+U1F). Experimental values were
taken from Ref. \cite{Audi2017}.} \label{82}
\end{figure}

Our investigated EE half-lives (GT and GT+U1F) for waiting point
isotones with $N$= 50 ($^{76}$Fe, $^{78}$Ni, $^{80}$Zn) is shown in
Figure.~\ref{50} which consists of three panels. The left and middle
panels compare our computed half-lives of allowed GT with GT
half-lives of Refs.~\cite{Moller1997,Lank03,Mol03,Pfe02}. In Ref
~\cite{Pfe02}, the KHF formula was implemented and two additional
types of QRPA calculation, namely QRPA-1 and QRPA-2, were performed.
KHF stands for Kratz-Hermann Formula which was developed by
Kratz and Herrmann \cite{Kratz73}. They applied the concept of the
$\beta$-strength function to the integral quantity of the
delayed-neutron emission probability and derived a simple
phenomenological expression for the calculation of neutron emission
probabilities and half-lives. For details of the model calculation
we refer to \cite{Pfe02}. The right panel compare the allowed
GT+U1F half-lives of our pn-QRPA model with theoretical
computations of allowed GT and forbidden contributions of
Refs.~\cite{Zhi13,Mol03,Bor05}. In Figure.~\ref{82}, we show a
similar comparison of our calculation for heavy $N$= 82 ($^{126}$Ru,
$^{128}$Pd, $^{130}$Cd) waiting point isotones. The left and middle
panels of Figure.~\ref{82} compare the pn-QRPA model
calculated allowed GT half-lives with various theoretical and
experimental results. The right panel compares the modified
half-lives with inclusion of forbidden contributions. All measured
half-lives were taken from \cite{Audi2017}. The present computed
results for half-life based on the pn-QRPA model for $N$=
50 and 82 waiting points are in good comparison with the measured
data. The addition of U1F transition resulted in a better comparison
of our calculated half-lives with the measured ones. In case
of N = 50, the parity changing transitions contribute significantly
to the decay rates and hence reduces the total half-lives
calculation effectively. For N = 82, the relative contribution of
U1F weak rates is comparatively smaller but still significant.

We employed the  nuclear deformation parameter ($\beta_2$) values
as a free parameter in our calculation. Our first set of
computation was based on the $\beta_2$ values taken from recent globally
updated calculation of FRDM \cite{Mol2012}, which we refer to as
pn-QRPA-FRDM. The second and third set of calculations
were based on values of $\beta_2$ computed from the RMF model using
density-dependent DD-ME2 and DD-PC1 interactions, respectively. We
refer to these calculations as pn-QRPA-RMF[DD-ME2] and
pn-QRPA-RMF[DD-PC1], respectively.  Table.~(\ref{Beta-RPA-RMF}) shows the three set of values of deformation parameter used in our calculation. We first investigated how the
computed EE rates changed by using different values of deformation.

In Tables.~(\ref{RPARMFME2},~\ref{RPARMFPC1}) we display the ratio
of calculated pn-QRPA-FRDM to pn-QRPA-RMF[DD-ME2] rates and
pn-QRPA-FRDM to pn-QRPA-RMF[DD-PC1] rates, respectively. The ratio
includes allowed GT and U1F rates separately. The ratios are
computed at $\rho Y_{e}$ = (10$^{4}$, 10$^{7}$, 10$^{11}$) and
$T_9$~=~(5, 15, 30), where $\rho Y_{e}$ and  $T_9$ represent the
stellar density and temperature in units of \emph{g/cm$^{3}$} and
$10^{9}$~K, respectively.
 Tables.~(\ref{RPARMFME2},~\ref{RPARMFPC1}) show that the  allowed GT pn-QRPA-FRDM rates are bigger by an order of magnitude than   pn-QRPA-RMF[DD-ME2] and pn-QRPA-RMF[DD-PC1] rates for $^{78}$Ni ($N$= 50) and  $N$= 82 waiting points.  For  $^{76}$Fe and $^{80}$Zn, the pn-QRPA-FRDM computed allowed GT rates are smaller by the same order of magnitude than the calculated rates by pn-QRPA-RMF[DD-ME2] and pn-QRPA-RMF[DD-PC1].  We note that for larger values of deformation parameters the  allowed GT rates are smaller. In case of $^{130}$Cd
and $^{126}$Ru, at high stellar density ($\rho Y_{e}$ = 10$^{11}$ \emph{g/cm$^{3}$}) and temperature ($T_9$~=~30 \emph{$GK$}) the ratio of allowed GT for  pn-QRPA-FRDM to pn-QRPA-RMF[DD-ME2] and pn-QRPA-RMF[DD-PC1] are an order of magnitude bigger. We further note that all three sets of weak rates increase with increasing temperature and decrease with increasing density values of the core for both GT and U1F transitions. With increasing temperature the number of parent states, that contribute to the total rates, increases. On the other hand, with increasing density the rates start to decrease
appreciably due to a decrease in the available phase space. 
We decided to proceed  with the investigation on the total weak rates (Eq.~(\ref{lb})) based only on
the pn-QRPA-FRDM model. The reason for this choice was that the FRDM predicts essentially a spherical shape for these closed-shell magic nuclei. We present our results for weak rates using the pn-QRPA-FRDM model in the succeeding paragraphs. 

In Tables.~(\ref{PC=EE=76Fe},~\ref{PC=EE=78Ni},~\ref{PC=EE=80Zn})
we present the computed allowed GT and U1F weak rates for $N=$
50 waiting point isotones ({$^{76}$Fe}, {$^{78}$Ni} and
{$^{80}$Zn}) using the pn-QRPA-FRDM model. Similarly in
Tables.~(\ref{PC=EE=126Ru},~\ref{PC=EE=128Pd},~\ref{PC=EE=130Cd})
we show the corresponding results for heavy $N=$ 82 waiting point
isotones ($^{126}$Ru, $^{128}$Pd and $^{130}$Cd). The computed rates
are presented at three different stellar density values of 10$^{4}$
~\emph{g/cm$^{3}$} (representing low-density region), 10$^{7}$
~\emph{g/cm$^{3}$} (representing intermediate-density region) and
10$^{11}$ ~\emph{g/cm$^{3}$} (representing high-density region). The
pn-QRPA computed PC and EE rates are shown in logarithmic (to base
10) scale in units of \emph{s$^{-1}$}.
We note that the computed
PC rates increase as the stellar
temperature rises. This makes sense as positrons are created only at high stellar temperatures to get captured. In case of
U1F rates one note that the contribution of PC rates are rather significant. Specially at $T_{9}$= 30, the computed U1F positron capture rates contribute
almost 90\% for most waiting point isotones. It is evident from
Table.~(\ref{PC=EE=76Fe}) to Table.~(\ref{PC=EE=130Cd}) that PC
rates must be taken into account, specially at high core temperatures, as they
 compete well with EE rates for most of the waiting point
isotones. Our findings are significant and emphasize that PC rates
of waiting point nuclei need to be taken into account in simulation codes, specially at high stellar
temperatures.

\begin{figure}[h!t]
\centering
\includegraphics[width=0.6\textwidth]{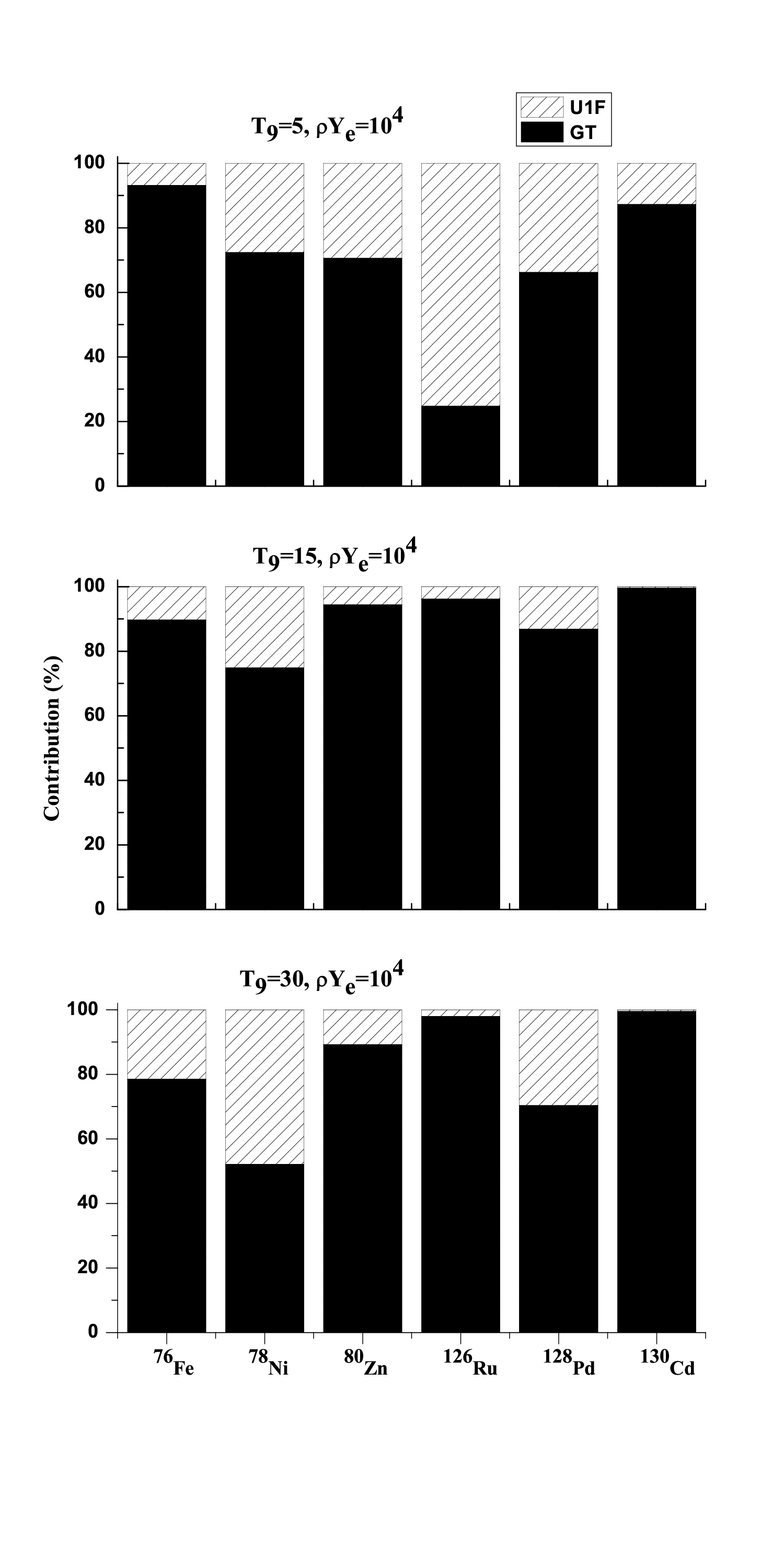}
\caption{Percentage contribution of allowed GT and U1F rates to
total electron emission $\lambda_{EE}$ rates. Stellar density
$\rho$Y$_{e}$ is stated in units of \emph{g/cm$^{3}$}, while
temperature T$_{9}$ is given in units of \emph{\rm 10$^{9}~K$}.}
\label{4}
\end{figure}

\begin{figure}[h!t]
\centering
\includegraphics[width=0.6\textwidth]{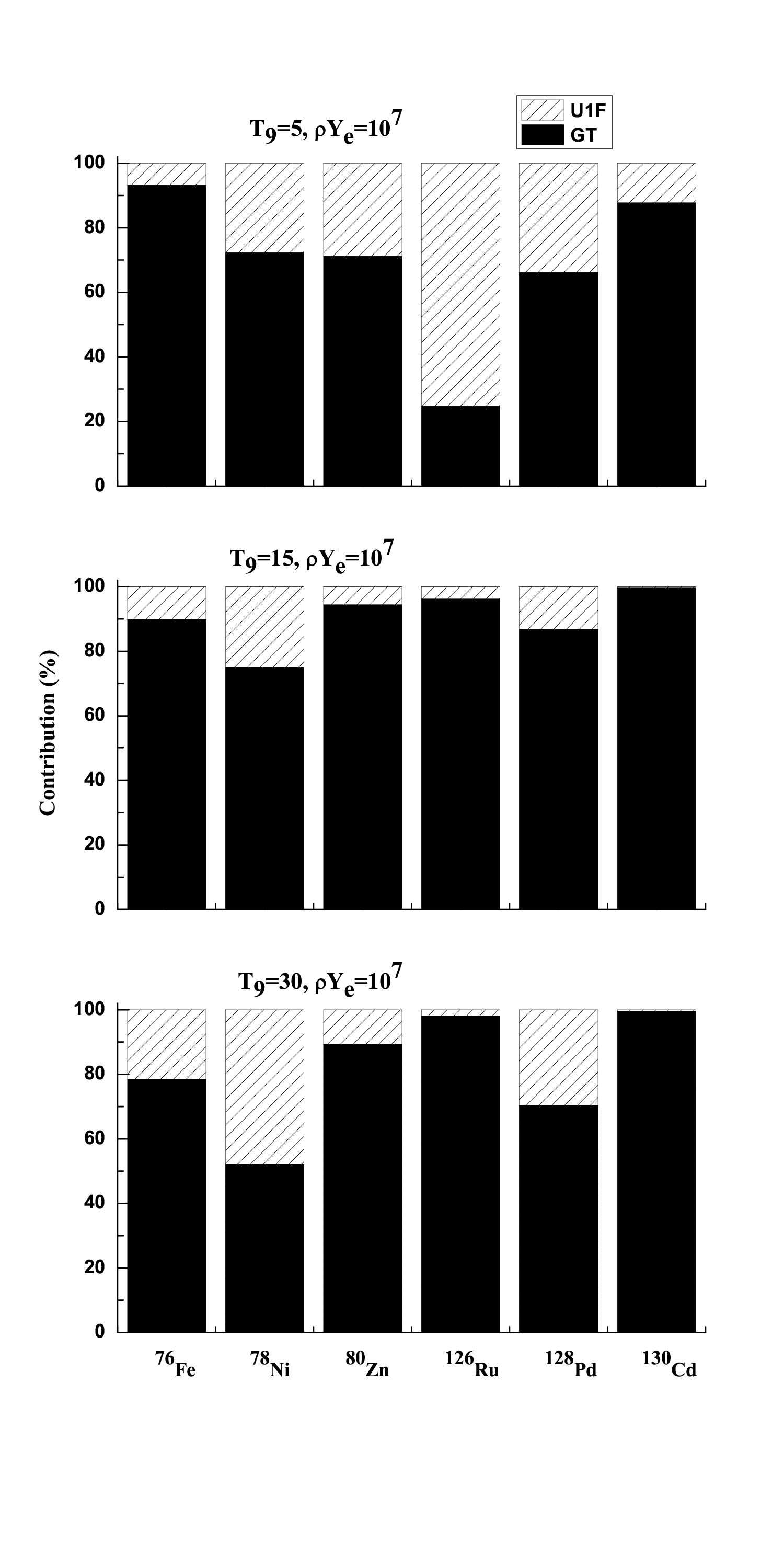}
\caption{Same as Figure.~\ref{4} but for stellar density 10$^{7}
g/cm^{3}$.} \label{7}
\end{figure}

\begin{figure}[h!t]
\centering
\includegraphics[width=0.6\textwidth]{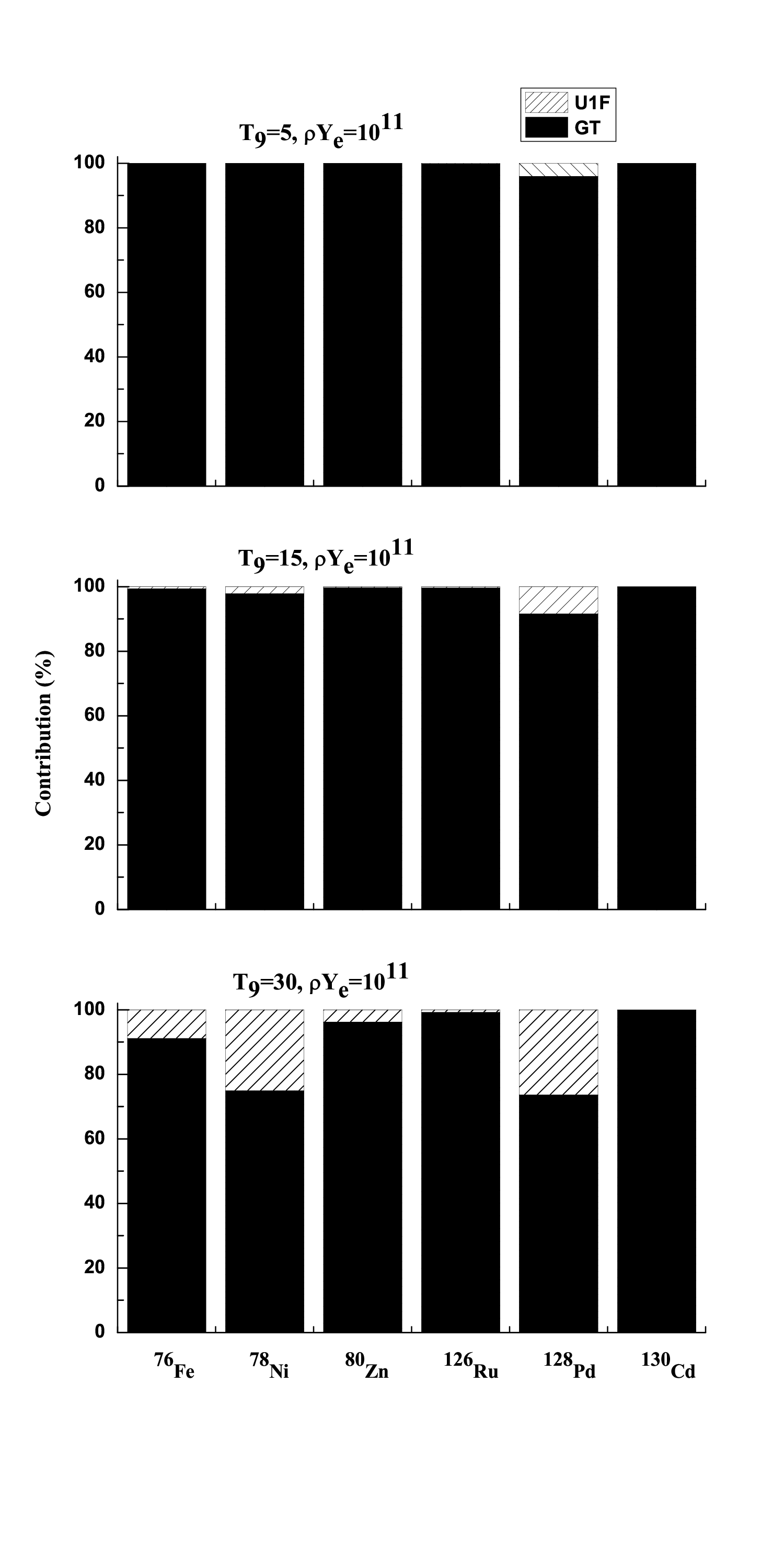}
\caption{Same as Figure.~\ref{4} but for stellar density 10$^{11}
g/cm^{3}$.} \label{11}
\end{figure}

We next investigate the contribution of U1F rates to the total EE rates using the pn-QRPA-FRDM model. In
Figs.~(\ref{4},~\ref{7},~\ref{11}), we depict the percentage
contribution of our computed allowed GT and U1F rates to total EE
rates. Fig.~\ref{4}, Fig.~\ref{7} and Fig.~\ref{11} are drawn at low ($\rho Y_{e}$= 10$^{4}$
\emph{g/cm$^{3}$}),
intermediate ($\rho Y_{e}$= 10$^{7}$
\emph{g/cm$^{3}$}) and high ($\rho Y_{e}$= 10$^{11}$
\emph{g/cm$^{3}$}) core density, respectively. Each figure consists of three panels depicting the
respective contribution of allowed GT and U1F rates at
temperatures $T_{9}$= (5, 15 and 30). We note
that the contribution of U1F rates reduces with increasing stellar
density. The U1F electron emission rates contribute reasonably well at low and
intermediate stellar densities.  We note
that for stellar temperatures ($T_{9} \leq$ 15) and high stellar
density values, almost all contribution to total EE rates comes from the
GT transitions. The EE rates (GT+U1F) on a fine grid temperature-density
scale, suitable for interpolation purposes, are available as ASCII
files and may be requested from the corresponding author.

%%%%%%%%%%%%%%%%%%%%%%%%%%%%%%%%%%%%%%%%%%%%%%%%%%%%%%%%%%%%%%%%%%%%%%%%%%%%%%%%%%%%%%%%%%%%%%%%%%%%%%%%%

%%%%%%%%%%%%%%%%%%%%%%%%%%%%%% Summary and Conclusion  %%%%%%%%%%%%%%%%%%%%%%%%%%%%%%%%%%%%%%%%%%%%%%%%%%%%%%%%%%%%%%%%%%%%%%%%%%%%%%%%%%

\section{Conclusion}

In the first investigation of the present study we have used the
axially deformed RMF model to calculate the nuclear ground state
properties for $N$ = 50 and 82 waiting point isotones. We presented
our computations of the ground-state binding energies, $S_{n}$,
$S_{p}$, $S_{2n}$ $S_{2p}$, charge, neutron-proton radii,
deformation parameters and quadrupole moments. Two different kinds
of density-dependent interactions, namely DD-ME2 and DD-PC1
interactions, were employed in the axially deformed RMF model for
the investigation of nuclear ground state properties. The calculated
binding energies were found to be in decent agreement with available
experimental data. The calculated PECs and PESs of the waiting point
isotones show that the ground-state shape of nuclei, studied in this
project, are almost spherical.

The second investigation of the present study employed the pn-QRPA model for calculation of weak rates under terrestrial and stellar conditions. EE and PC rates for a wide range of stellar temperature
and density were calculated for allowed GT and U1F transitions.
Our half-life results were in good agreement with the measured data.
The addition of U1F rates appreciably reduced the
total calculated half-lives. Our half-life calculation may
further be improved by incorporation of non-unique forbidden transitions
(rank 0 and 1). We plan to compute non-unique forbidden transitions as a future assignment. We investigated the role of deformation parameter, computed in the RMF model, in our pn-QRPA calculation  and concluded that the rates changed by up to an order of magnitude or more by changing the deformation parameters.
We further
investigated the competition between PC and EE rates. The PC rates on waiting point isotones were found to
be significant especially at high core temperatures and low stellar densities.  We noted that contribution of U1F
rates are significant to the total EE rates for $N$ = 50 and 82 waiting point isotones. The results presented
here might prove useful  for the modeling of
presupernova evolution of massive stars.
%%%%%%%%%%%%%%%%%%%%%%%%%%%%%%%%%%%%%%%%%%%%%%%%%%%%%%%%%%%%%%%%%%%%%%%%%%%%%%

%%%%%%%%%%%%%%%%%%%%%%%%%%%%%%%%%%%%%%%%%%%%%%%%%%%%%%%%%%%%%%%%%%%%%%%%%%%%%%%%%%%%%%%%%%%%%%%%%%%%%%%%%%%%%%%%%%%%%%%%%%%%%%%%%%%%%%%%%%%%
%%%%%%%%%%%%%%%%%%%%%%%%    RMF Figures   %%%%%%%%%%%%%%%%%%%%%%%%%%%%%%%%%%%%%%%%%%%%%%%
\clearpage

\clearpage

\clearpage

\clearpage

\clearpage

\clearpage

%%%%%%%%%%%%%%%%%%%%%%%%%%%%%%%%%%%%%%%%%%%%%%%%%%%%%%%%%%%%%%%%%%%%%%%%%%%%%%%%%%%%%%%%%%%%%%%%%%%%%%%%%%%%%

%%%%%%%%%%%%%%%%%%%%%%%%%%%%%%%%%%%%%%%%%%%%%%%%%%%%%%%%%%%%%%%%%%%%%%%%%%%%%%%%%%%%%%%%%%%%%%%%%%

%%%%%%%%%%%%%%%%%%%%%%%%%%%%%%%%%%%%%%%%%%%%%%%%%%%   Tables   %%%%%%%%%%%%%%%%%%%%%%%%%%%%%%%%%%%%%%%%%%%%%%%%%%%%%%%%%%%%%%%%%%%%%%%
\clearpage
\begin{table*}
\centering\scriptsize\caption{Calculated charge radii of
$^{76}$Fe, $^{78}$Ni, $^{80}$Zn, $^{126}$Ru, $^{128}$Pd and
$^{130}$Cd using the RMF model with DD-ME2 and DD-PC1
interactions in comparison with HFB and empirical results. All units
are in $fm$.} \label{charge}

%\centering \footnotesize\setlength{\tabcolsep}{1.5pt}
\begin{tabular}{c|c|c|c|c|c}
\hline Nucleus & RMF[DD-ME2] & RMF[DD-PC1] &
RMF[NL3*]~\cite{bayram2013a} & HFB[SLy4]~\cite{stoitsov2003} &
Empirical~\cite{bayram2013c} \cr

\hline

$^{76}$Fe & $3.924$ & $3.936$ & $3.787$ & $3.956$ & $3.818$  \\
\hline

$^{78}$Ni & $3.964$ & $3.979$ & $3.979$ & $4.004$ & $3.906$ \\
\hline

$^{80}$Zn & $4.023$ & $4.039$ & $4.038$ & $4.055$ & $3.991$ \\
\hline

$^{126}$Ru & $4.630$ & $4.629$ & $4.646$ & $4.639$ & $4.500$ \\
\hline

$^{128}$Pd & $4.661$ & $4.663$ & $4.676$ & $4.674$ & $4.564$ \\
\hline

$^{130}$Cd & $4.691$ & $4.695$ & $4.705$ & $4.706$ & $4.626$ \\
\hline
\end{tabular}
\end{table*}

%\clearpage
\begin{table*}
\centering\scriptsize\caption{Calculated ground-state
properties of $^{76}$Fe, $^{78}$Ni, $^{80}$Zn, $^{126}$Ru,
$^{128}$Pd and $^{130}$Cd using the RMF model with DD-ME2
interaction.} \label{ddme2}

%\centering \footnotesize\setlength{\tabcolsep}{1.5pt}
\begin{tabular}{c|c|c|c|c|c|c|c|c|c|c}
\hline Nuclei & BE/A & S$_{n}$ & S$_{p}$ & S$_{2n}$ & S$_{2p}$ &
r$_{n}$ & r$_{p}$ & r$_{c}$ & $\beta_{2}$ & Q$_{T}$ \cr
  & [MeV] & [MeV] & [MeV] & [MeV] & [MeV] & [fm]  & [fm] & [fm] &  & [barn] \cr
\hline

$^{76}$Fe & $7.877$ & $2.288$ & $21.963$ & $4.554$ & $44.674$ &
$4.219$ & $3.841$ & $3.924$ & 0.0001 & $0.001$ \\
 \hline

$^{78}$Ni &$8.211$ & $4.379$ & $20.596$ & $8.856$ & $41.878$ &
$4.198$ &
$3.880$ & $3.961$ & 0.0004 & $0.006$ \\
 \hline

$^{80}$Zn & $8.397$ &$5.403$ & $15.302$ & $10.065$ & $31.265$ &
$4.210$ & $3.942$ & $4.023$ & 0.0012 & $0.020$ \\
\hline

$^{126}$Ru & $7.974$ & $4.040$ & $18.280$ & $8.103$ & $37.093$ &
$4.922$ & $4.561$ & $4.630$ & $-0.0014$ & $-0.048$ \\
 \hline

$^{128}$Pd & $8.123$ & $4.875$ & $17.244$ & $9.683$ & $35.002$ &
$4.915$ & $4.592$ & $4.661$ & $-0.0015$ & $-0.054$ \\
 \hline

$^{130}$Cd & $8.251$ & $5.703$ & $16.232$ & $11.476$ & $32.968$ &
$4.909$ & $4.622$ & $4.691$ & $-0.0015$ & $-0.054$ \\
\hline
\end{tabular}
\end{table*}

%%%%%%%%%%%%%%%%%%%%%%%%%%%%%%%%%%%%%%%%%%%%%%%%%%%%%%%%%%%%%%%%%%%%%%%

%\clearpage
\begin{table*}
\centering\scriptsize\caption{Same as Table~\ref{ddme2} but for RMF
model with DD-PC1 interaction.} \label{ddpc1}

%\centering \footnotesize\setlength{\tabcolsep}{1.5pt}
\begin{tabular}{c|c|c|c|c|c|c|c|c|c|c}
\hline
Nuclei & BE/A & S$_{n}$ & S$_{p}$ & S$_{2n}$ & S$_{2p}$ &
r$_{n}$ & r$_{p}$ & r$_{c}$ & $\beta_{2}$ & Q$_{T}$ \cr
  & [MeV] & [MeV] & [MeV] & [MeV] & [MeV] & [fm]  & [fm] & [fm] &  & [barn] \\
\hline

$^{76}$Fe & $7.970$ & $3.377$ & $20.947$ & $5.879$ & $42.581$ &
$4.240$ & $3.854$ & $3.936$ & $-0.0001$ & $-0.002$ \\
\hline

$^{78}$Ni & $8.277$ & $4.564$ & $19.545$ & $9.259$ & $39.858$ &
$4.225$ & $3.898$ & $3.979$ & $-0.0011$ & $-0.017$ \\
 \hline

$^{80}$Zn & $8.447$ & $5.619$ & $14.769$ & $10.625$ & $30.201$ &
$4.236$ & $3.959$ & $4.039$ & $-0.0058$ & $-0.095$ \\
 \hline

$^{126}$Ru & $8.040$ & $4.425$ & $17.613$ & $8.932$ & $35.755$ &
$4.930$ & $4.559$ & $4.629$  & $-0.0008$ & $-0.026$ \\
 \hline

$^{128}$Pd & $8.178$ & $5.189$ & $16.584$ & $10.465$ & $33.679$ &
$4.926$ & $4.594$ & $4.663$ & $-0.0011$ & $-0.039$ \\
 \hline

$^{130}$Cd & $8.295$ & $5.950$ & $15.572$ & $11.991$ & $31.650$ &
$4.923$ & $4.627$ & $4.695$ & $-0.0013$ & $-0.047$ \\
 \hline
\end{tabular}
\end{table*}
%%%%%%%%%%%%%%%%%%%%%%%%%%%%%%%%%%%%%%%%%%%%%%%%%%%%%%%%%%%%%%%%%%%%%%%%%%%%%%%%%%%%%%%%%%%%%%%%%%%%%%%%%%%%%%%%%%%%%%%%%%%%%%%%%%%%%%%
\begin{table*}
\centering\scriptsize\caption{Calculated deformation parameters
($\beta_{2}$) used by pn-QRPA calculation from FRDM \cite{Mol2012}
and those deduced from the RMF model with DDME2 and DDPC1
interactions.} \label{Beta-RPA-RMF}

%\centering \footnotesize\setlength{\tabcolsep}{1.5pt}
\begin{tabular}{c|c|c|c}
\hline Nuclei & $\beta_{2}$(FRDM) & $\beta_{2}$(RMF[DD-ME2]) &
$\beta_{2}$(RMF[DD-PC1]) \cr \hline

$^{76}$Fe       & -0.01100 & 0.00010  & -0.00010 \\
\hline
$^{78}$Ni       & 0.00000  & 0.00040  & -0.00110 \\
\hline
$^{80}$Zn        & -0.01000 & 0.00120  & -0.00580 \\
\hline
$^{126}$Ru      & 0.00000  & -0.00140 & -0.00080 \\
\hline
$^{128}$Pd      & 0.00000  & -0.00150 & -0.00110 \\
\hline
$^{130}$Cd     & 0.00000  & -0.00150 & -0.00130 \\
\hline
\end{tabular}
\end{table*}

%%%%%%%%%%%%%%%%%%%%%%%%%%%%%%%%%%%%%%%%%%%%%%%%%%%%%%%%%%%%%%%%%%%%%%%%%%%%%%%%%%%%%%%%%%%%%%%%%%%%%%%%%%%%%%%%%%%%%%%%%%
\begin{table*}
\centering\scriptsize\caption{Ratio of computed (allowed GT and U1F
electron emission ($\lambda_{EE}$)) pn-QRPA-FRDM to
pn-QRPA-RMF[DD-ME2] rates as a function of stellar temperature and
density. Stellar temperature (T$_{9}$) and density ($\rho$Y$_{e}$)
are shown in units of {10$^{9}$~\it K} and \emph{g/cm$^{3}$},
respectively.} \label{RPARMFME2}
%\centering
%\footnotesize\setlength{\tabcolsep}{1.5pt}
\begin{tabular}{c|c|c|c|c|c|c|c}
\hline
Nucleus & T$_{9}$ & \multicolumn{3}{c|} { Ratio [Allowed GT]} & \multicolumn{3}{c} {Ratio [U1F]} \\
\cline{3-8} & & $\rho$$\it Y_{e}$=10$^{4}$ & $\rho$$\it
Y_{e}$=10$^{7}$ & $\rho$$\it Y_{e}$=10$^{11}$ & $\rho$$\it
Y_{e}$=10$^{4}$ & $\rho$$\it Y_{e}$=10$^{7}$ & $\rho$$\it
Y_{e}$=10$^{11}$ \\
\hline
                      & 5  & 5.62E-01 & 5.62E-01 & 5.66E-01 & 1.38E+00 & 1.38E+00 & 3.50E+00 \\
$^{76}$Fe             & 15 & 2.40E-01 & 2.40E-01 & 2.42E-01 & 1.66E+00 & 1.66E+00 & 3.78E+00 \\
                      & 30 & 1.21E-01 & 1.21E-01 & 1.17E-01 & 2.09E+00 & 2.09E+00 & 4.26E+00 \\
\hline
                      & 5  & 1.00E+00 & 1.00E+00 & 1.32E+00 & 9.89E-01 & 9.89E-01 & 8.53E-01 \\
$^{78}$Ni             & 15 & 2.26E+00 & 2.26E+00 & 4.24E+00 & 8.47E-01 & 8.47E-01 & 7.73E-01 \\
                      & 30 & 4.11E+00 & 4.11E+00 & 5.38E+00 & 6.32E-01 & 6.32E-01 & 6.00E-01 \\
\hline
                      & 5  & 9.44E-01 & 9.44E-01 & 9.89E-01 & 1.02E+00 & 1.03E+00 & 1.04E+00 \\
$^{80}$Zn             & 15 & 7.69E-01 & 7.71E-01 & 9.66E-01 & 1.03E+00 & 1.04E+00 & 1.06E+00 \\
                      & 30 & 6.18E-01 & 6.19E-01 & 6.95E-01 & 1.09E+00 & 1.09E+00 & 1.09E+00 \\
\hline
                      & 5  & 1.16E+00 & 1.15E+00 & 3.30E+00 & 9.55E-01 & 9.62E-01 & 2.05E-01 \\
$^{126}$Ru            & 15 & 6.43E+00 & 6.43E+00 & 3.84E+00 & 9.51E-01 & 9.31E-01 & 2.04E-01 \\
                      & 30 & 7.64E+00 & 7.62E+00 & 6.14E+00 & 8.59E-01 & 9.18E-01 & 1.99E-01 \\
\hline
                      & 5  & 2.31E+00 & 2.32E+00 & 1.11E+00 & 9.42E-01 & 9.35E-01 & 8.20E-01 \\
$^{128}$Pd            & 15 & 2.52E+00 & 2.52E+00 & 1.46E+00 & 8.89E-01 & 8.89E-01 & 5.04E-01 \\
                      & 30 & 2.56E+00 & 2.56E+00 & 1.51E+00 & 4.66E-01 & 4.67E-01 & 3.90E-01 \\
\hline
                      & 5  & 1.22E+00 & 1.23E+00 & 5.92E+01 & 9.82E-01 & 9.84E-01 & 8.09E-01 \\
$^{130}$Cd            & 15 & 1.96E+01 & 1.97E+01 & 7.01E+01 & 3.31E-01 & 3.31E-01 & 2.94E-01 \\
                      & 30 & 2.12E+01 & 2.12E+01 & 3.73E+01 & 1.94E-01 & 1.94E-01 & 1.83E-01 \\
\hline
%\hline
\end{tabular}
\end{table*}

%%%%%%%%%%%%%%%%%%%%%%%%%%%%%%%%%%%%%%%%%%%%%%%%%%%%%%%%%%%%%%%%%%%%%%%%%%%%%%%%%%%%%%%%%%%%%%%%%%%%%%%%%%%%%

\begin{table*}
\centering\scriptsize\caption{Ratio of computed (allowed GT and U1F
electron emission ($\lambda_{EE}$)) pn-QRPA-FRDM to
pn-QRPA-RMF[DD-PC1] rates as a function of stellar temperature and
density. Stellar temperature (T$_{9}$) and density ($\rho$Y$_{e}$)
are shown in units of {10$^{9}$~\it K} and \emph{g/cm$^{3}$},
respectively.} \label{RPARMFPC1}
%\centering
%\footnotesize\setlength{\tabcolsep}{1.5pt}
\begin{tabular}{c|c|c|c|c|c|c|c}
\hline Nucleus & T$_{9}$ &
\multicolumn{3}{c|}{ Ratio [Allowed GT]} &\multicolumn{3}{c}{Ratio [U1F]} \\
\cline{3-8} & & $\rho$$\it Y_{e}$=10$^{4}$ & $\rho$$\it
Y_{e}$=10$^{7}$ & $\rho$$\it Y_{e}$=10$^{11}$ & $\rho$$\it
Y_{e}$=10$^{4}$ & $\rho$$\it Y_{e}$=10$^{7}$ & $\rho$$\it
Y_{e}$=10$^{11}$ \\
\hline
                    & 5  & 5.62E-01 & 5.62E-01 & 5.66E-01 & 1.38E+00 & 1.38E+00 & 3.53E+00 \\
$^{76}$Fe             & 15 & 2.40E-01 & 2.40E-01 & 2.42E-01 & 1.66E+00 & 1.66E+00 & 3.78E+00 \\
                      & 30 & 1.21E-01 & 1.21E-01 & 1.17E-01 & 2.09E+00 & 2.09E+00 & 4.26E+00 \\
\hline
                      & 5  & 1.01E+00 & 1.01E+00 & 1.16E+00 & 9.89E-01 & 9.89E-01 & 8.53E-01 \\
$^{78}$Ni             & 15 & 2.33E+00 & 2.33E+00 & 4.23E+00 & 8.05E-01 & 8.05E-01 & 7.53E-01 \\
                      & 30 & 5.05E+00 & 5.05E+00 & 6.64E+00 & 6.12E-01 & 6.12E-01 & 5.83E-01 \\
\hline
                      & 5  & 9.77E-01 & 9.77E-01 & 7.33E-01 & 1.01E+00 & 1.01E+00 & 1.05E+00 \\
$^{80}$Zn             & 15 & 8.34E-01 & 8.30E-01 & 6.14E-01 & 1.01E+00 & 1.01E+00 & 1.07E+00 \\
                      & 30 & 7.18E-01 & 7.18E-01 & 6.04E-01 & 1.02E+00 & 1.03E+00 & 1.15E+00 \\
\hline
                      & 5  & 1.13E+00 & 1.12E+00 & 9.82E+00 & 9.62E-01 & 9.62E-01 & 8.17E-01 \\
$^{126}$Ru            & 15 & 9.20E+00 & 9.20E+00 & 9.31E+00 & 9.51E-01 & 9.57E-01 & 5.75E-01 \\
                      & 30 & 9.40E+00 & 9.40E+00 & 1.01E+01 & 9.40E-01 & 9.20E-01 & 2.56E-01 \\
\hline
                      & 5  & 2.38E+00 & 2.39E+00 & 1.82E+00 & 9.46E-01 & 9.38E-01 & 8.20E-01 \\
$^{128}$Pd            & 15 & 2.40E+00 & 2.40E+00 & 1.96E+00 & 8.69E-01 & 8.69E-01 & 5.01E-01 \\
                      & 30 & 2.46E+00 & 2.52E+00 & 2.03E+00 & 4.49E-01 & 4.49E-01 & 3.94E-01 \\
\hline
                      & 5  & 1.22E+00 & 1.22E+00 & 7.13E+01 & 9.95E-01 & 9.95E-01 & 8.45E-01 \\
$^{130}$Cd            & 15 & 2.03E+01 & 2.04E+01 & 7.48E+01 & 3.28E-01 & 3.28E-01 & 2.88E-01 \\
                      & 30 & 2.15E+01 & 2.15E+01 & 3.84E+01 & 1.86E-01 & 1.86E-01 & 1.74E-01 \\
\hline
%\hline
\end{tabular}
\end{table*}
%%%%%%%%%%%%%%%%%%%%%%%%%%%%%%%%%%%%%%%%%%%%%%%%%%%%%%%%%%%%%%%%%%%%%%%%%%%%%%%%%%%%%%%%%%%%%%%%%%%%%%%%%%%%%%%%%%
%%%%%%%%%%%%%%%%%%%%%%%%%%%%%%%%%%%%%%%%%%%%%%%%%%%%%%%%%%%%%%%%%%%%%%%%%%%%%%%%%%%%%%%%%%%%%%%%%%%%%%%%%%%%%%%

\begin{table*}
\centering\scriptsize\caption{Computed allowed GT and U1F positron
capture ($\lambda_{PC}$) and electron emission ($\lambda_{EE}$)
rates on {$^{76}$Fe}. \%$_{PC}$ and \%$_{EE}$ represent the
percentage contribution of $\lambda_{PC}$ and $\lambda_{EE}$ rates,
respectively, to the total rate. $\rho$Y$_{e}$ and T$_{9}$ are
stated in units of \emph{g/cm$^{3}$} and {10$^{9}$~\it K},
respectively. Rates
are stated in logarithmic (to base 10) scale in units of
\emph{s$^{-1}$}.} \label{PC=EE=76Fe}
    \begin{tabular}{c|c|cccc|cccc}
\hline
$\rho$Y$_{e}$ & T$_{9}$ & \multicolumn{4}{c|}{$^{76}$Fe [Allowed GT]} & \multicolumn{4}{c}{$^{76}$Fe [U1F]} \\
\cline{3-10}& & $\lambda_{PC}$(s$^{-1}$)& $\lambda_{EE}$(s$^{-1}$)& \%$_{PC}$&\%$_{EE}$ & $\lambda_{PC}$(s$^{-1}$) & $\lambda_{EE}$(s$^{-1}$)& \%$_{PC}$& \%$_{EE}$ \\
\hline
                       & 5  & -1.62  & 1.46   & 0.08  & 99.92  & -1.21  & 1.53   & 0.18  & 99.82  \\
10$^{4}$              & 15 & 0.42   & 1.75   & 4.43  & 95.57  & 1.48   & 1.93   & 26.01 & 73.99  \\
                      & 30 & 1.80   & 1.96   & 41.00 & 59.00  & 3.62   & 2.43   & 93.88 & 6.12   \\
\hline
                      & 5  & -2.30  & 1.45   & 0.02  & 99.98  & -1.91  & 1.53   & 0.04  & 99.96  \\
10$^{7}$              & 15 & 0.39   & 1.75   & 4.17  & 95.83  & 1.45   & 1.93   & 24.83 & 75.17  \\
                      & 30 & 1.80   & 1.96   & 40.84 & 59.16  & 3.61   & 2.43   & 93.84 & 6.16   \\
\hline
                      & 5  & -25.70 & -10.53 & 0.00  & 100.00 & -25.30 & -14.34 & 0.00  & 100.00 \\
10$^{11}$             & 15 & -7.52  & -2.69  & 0.00  & 100.00 & -6.47  & -4.01  & 0.34  & 99.66  \\
                      & 30 & -2.04  & -0.41  & 2.26  & 97.74  & -0.24  & -0.46  & 62.56 & 37.44  \\
 \hline
\end{tabular}
\end{table*}
%%%%%%%%%%%%%%%%%%%%%%%%%%%%%%%%%%%%%%%%%%%%%%%%%%%%%%%%%%%%%%%%%%%%%%%%%%%%%%%%%%%%%%%%%%%%%%%%%%%%%%%%%%%%%%%%%%%
\begin{table*}
    \centering\scriptsize\caption{Same as Table.~\ref{PC=EE=76Fe}
but for {$^{78}$Ni}.}
\label{PC=EE=78Ni}
\begin{tabular}{c|c|cccc|cccc}
        \hline
        $\rho$Y$_{e}$ & T$_{9}$ & \multicolumn{4}{c|}{$^{78}$Ni [Allowed GT]} & \multicolumn{4}{c}{$^{78}$Ni [U1F]} \\
        \cline{3-10}& & $\lambda_{PC}$(s$^{-1}$)& $\lambda_{EE}$(s$^{-1}$)& \%$_{PC}$&\%$_{EE}$) & $\lambda_{PC}$(s$^{-1}$) & $\lambda_{EE}$(s$^{-1}$)& \%$_{PC}$& \%$_{EE}$ \\
        \hline
                      & 5  & -1.99  & 0.53   & 0.30  & 99.70  & -1.45  & 0.81   & 0.54  & 99.46  \\
10$^{4}$              & 15 & 0.52   & 1.58   & 8.05  & 91.95  & 1.22   & 1.07   & 58.49 & 41.51  \\
                      & 30 & 2.28   & 2.26   & 51.55 & 48.45  & 3.48   & 1.45   & 99.06 & 0.94   \\
 \hline
                      & 5  & -2.67  & 0.52   & 0.06  & 99.94  & -2.15  & 0.81   & 0.11  & 99.89  \\
10$^{7}$              & 15 & 0.50   & 1.58   & 7.61  & 92.39  & 1.19   & 1.07   & 57.03 & 42.97  \\
                      & 30 & 2.28   & 2.26   & 51.44 & 48.56  & 3.47   & 1.45   & 99.06 & 0.94   \\
 \hline
                      & 5  & -26.06 & -15.54 & 0.00  & 100.00 & -25.54 & -17.95 & 0.00  & 100.00 \\
10$^{11}$              & 15 & -7.41  & -3.43  & 0.01  & 99.99  & -6.73  & -5.39  & 4.35  & 95.65  \\
                      & 30 & -1.56  & -0.26  & 4.79  & 95.21  & -0.38  & -1.64  & 94.78 & 5.22   \\
 \hline
    \end{tabular}
\end{table*}

%%%%%%%%%%%%%%%%%%%%%%%%%%%%%%%%%%%%%%%%%%%%%%%%%%%%%%%%%%%%%%%%%%%%%%%%%%%%%%%%%%%%%%%%%%%%%%%%%%%%%%%%%%%%%%%%%%
\begin{table*}
\centering\scriptsize\caption{Same as Table.~\ref{PC=EE=76Fe}
but for {$^{80}$Zn}.}
\label{PC=EE=80Zn}
    \begin{tabular}{c|c|cccc|cccc}
\hline
$\rho$Y$_{e}$ & T$_{9}$ & \multicolumn{4}{c|}{$^{80}$Zn [Allowed GT]} & \multicolumn{4}{c}{$^{80}$Zn [U1F]} \\
\cline{3-10}& & $\lambda_{PC}$(\it s$^{-1}$)& $\lambda_{EE}$(\it s$^{-1}$)& \%$_{PC}$&\%$_{EE}$ & $\lambda_{PC}$(\it s$^{-1}$) & $\lambda_{EE}$(\it s$^{-1}$)& \%$_{PC}$& \%$_{EE}$ \\
\hline
                      & 5  & -2.21  & -0.09  & 0.75  & 99.25  & -1.73  & -0.47  & 5.24  & 94.76  \\
10$^{4}$              & 15 & 0.43   & 1.33   & 11.09 & 88.91  & 1.22   & -0.03  & 94.58 & 5.42   \\
                      & 30 & 1.83   & 1.53   & 66.61 & 33.39  & 3.55   & 0.46   & 99.92 & 0.08   \\
\hline
                      & 5  & -2.90  & -0.10  & 0.16  & 99.84  & -2.42  & -0.49  & 1.17  & 98.83  \\
10$^{7}$              & 15 & 0.40   & 1.33   & 10.51 & 89.49  & 1.19   & -0.03  & 94.29 & 5.71   \\
                      & 30 & 1.82   & 1.53   & 66.46 & 33.54  & 3.55   & 0.46   & 99.92 & 0.08   \\
\hline
                      & 5  & -26.29 & -16.74 & 0.00  & 100.00 & -25.81 & -21.27 & 0.00  & 100.00 \\
10$^{11}$             & 15 & -7.51  & -4.32  & 0.06  & 99.94  & -6.74  & -7.02  & 65.74 & 34.26  \\
                      & 30 & -2.02  & -1.25  & 14.57 & 85.43  & -0.31  & -2.83  & 99.70 & 0.30  \\
\hline
\end{tabular}
\end{table*}
%%%%%%%%%%%%%%%%%%%%%%%%%%%%%%%%%%%%%%%%%%%%%%%%%%%%%%%%%%%%%%%%%%%%%%%%%%%%%%%%%%%%%%%%%%%%%%%%%%%%%%%%%
\begin{table*}
\centering\scriptsize\caption{Same as Table.~\ref{PC=EE=76Fe}
but for {$^{126}$Ru}.}
\label{PC=EE=126Ru}
    \begin{tabular}{c|c|cccc|cccc}
\hline
$\rho$Y$_{e}$ & T$_{9}$ & \multicolumn{4}{c|}{$^{126}$Ru [Allowed GT]}  & \multicolumn{4}{c}{$^{126}$Ru [U1F]} \\
\cline{3-10}& & $\lambda_{PC}$(\it s$^{-1}$)& $\lambda_{EE}$(\it s$^{-1}$)& \%$_{PC}$&\%$_{EE}$ & $\lambda_{PC}$(\it s$^{-1}$) & $\lambda_{EE}$(\it s$^{-1}$)& \%$_{PC}$& \%$_{EE}$ \\
\hline
                      & 5  & -1.98  & 1.08   & 0.09  & 99.91  & -1.10  & 2.14   & 0.06  & 99.94  \\
10$^{4}$              & 15 & 1.34   & 3.02   & 2.07  & 97.93  & 1.64   & 2.22   & 20.98 & 79.02  \\
                      & 30 & 2.77   & 3.27   & 23.90 & 76.10  & 3.63   & 2.48   & 93.30 & 6.70   \\
\hline
                      & 5  & -2.67  & 1.07   & 0.02  & 99.98  & -1.80  & 2.14   & 0.01  & 99.99  \\
10$^{7}$              & 15 & 1.32   & 3.02   & 1.95  & 98.05  & 1.62   & 2.22   & 20.00 & 80.00  \\
                      & 30 & 2.76   & 3.27   & 23.82 & 76.18  & 3.62   & 2.48   & 93.26 & 6.74   \\
 \hline
                      & 5  & -26.06 & -12.07 & 0.00  & 100.00 & -25.19 & -14.59 & 0.00  & 100.00 \\
10$^{11}$             & 15 & -6.59  & -1.89  & 0.00  & 100.00 & -6.31  & -3.76  & 0.28  & 99.72  \\
                      & 30 & -1.08  & 0.77   & 1.41  & 98.59  & -0.23  & -0.50  & 65.22 & 34.78   \\
 \hline
\end{tabular}
\end{table*}
%%%%%%%%%%%%%%%%%%%%%%%%%%%%%%%%%%%%%%%%%%%%%%%%%%%%%%%%%%%%%%%%%%%%%%%%%%%%%%%%%%%%%%%%%%%%%%%%%%%%%%%%%%%%%%
\begin{table*}
    \centering\scriptsize\caption{Same as Table.~\ref{PC=EE=76Fe}
        but for {$^{128}$Pd}.}
    \label{PC=EE=128Pd}
    \begin{tabular}{c|c|cccc|cccc}
        \hline
        $\rho$Y$_{e}$ & T$_{9}$ & \multicolumn{4}{c|}{$^{128}$Pd [Allowed GT]}  & \multicolumn{4}{c}{$^{128}$Pd [U1F]} \\
        \cline{3-10}& & $\lambda_{PC}$(\it s$^{-1}$)& $\lambda_{EE}$(\it s$^{-1}$)& \%$_{PC}$&\%$_{EE}$ & $\lambda_{PC}$(\it s$^{-1}$) & $\lambda_{EE}$(\it s$^{-1}$)& \%$_{PC}$& \%$_{EE}$ \\
\hline
                     & 5  & -1.40  & 1.77   & 0.07  & 99.93  & -1.85  & 1.48   & 0.05  & 99.95  \\
10$^{4}$              & 15 & 1.11   & 2.48   & 4.08  & 95.92  & 0.86   & 1.66   & 13.63 & 86.37  \\
                      & 30 & 2.28   & 2.55   & 34.94 & 65.06  & 2.78   & 1.94   & 87.37 & 12.63  \\
\hline
                      & 5  & -2.08  & 1.77   & 0.01  & 99.99  & -2.55  & 1.47   & 0.01  & 99.99  \\
10$^{7}$              & 15 & 1.08   & 2.48   & 3.85  & 96.15  & 0.83   & 1.66   & 12.91 & 87.09  \\
                      & 30 & 2.28   & 2.55   & 34.73 & 65.27  & 2.77   & 1.94   & 87.27 & 12.73  \\
\hline
                      & 5  & -25.47 & -14.38 & 0.00  & 100.00 & -25.94 & -15.76 & 0.00  & 100.00 \\
10$^{11}$             & 15 & -6.83  & -3.29  & 0.03  & 99.97  & -7.09  & -4.33  & 0.17  & 99.83  \\
                      & 30 & -1.57  & -0.51  & 8.01  & 91.99  & -1.08  & -0.95  & 42.63 &   57.37\\
\hline
    \end{tabular}
\end{table*}

%%%%%%%%%%%%%%%%%%%%%%%%%%%%%%%%%%%%%%%%%%%%%%%%%%%%%%%%%%%%%%%%%%%%%%%%%%%%%%%%%%%%%%%%%%%%%%%%%%%%%%%%%
\begin{table*}
\centering\scriptsize\caption{Same as Table.~\ref{PC=EE=76Fe}
but for {$^{130}$Cd}.}
\label{PC=EE=130Cd}
    \begin{tabular}{c|c|cccc|cccc}
\hline
$\rho$Y$_{e}$ & T$_{9}$ & \multicolumn{4}{c|}{$^{130}$Cd [Allowed GT]} & \multicolumn{4}{c}{$^{130}$Cd [U1F]} \\
\cline{3-10}& & $\lambda_{PC}$(\it s$^{-1}$)& $\lambda_{EE}$(\it s$^{-1}$)& \%$_{PC}$&\%$_{EE}$ & $\lambda_{PC}$(\it s$^{-1}$) & $\lambda_{EE}$(\it s$^{-1}$)& \%$_{PC}$& \%$_{EE}$ \\
\hline
                      & 5  & -2.12  & 0.62   & 0.18  & 99.82  & -2.08  & -0.57  & 3.01  & 96.99 \\
10$^{4}$              & 15 & 1.03   & 2.47   & 3.48  & 96.52  & 0.90   & -0.28  & 93.82 & 6.18  \\
                      & 30 & 2.47   & 2.67   & 38.69 & 61.31  & 3.00   & 0.02   & 99.90 & 0.10  \\
\hline
                      & 5  & -2.81  & 0.61   & 0.04  & 99.96  & -2.77  & -0.60  & 0.67  & 99.33 \\
10$^{7}$              & 15 & 1.00   & 2.47   & 3.28  & 96.72  & 0.87   & -0.29  & 93.50 & 6.50  \\
                      & 30 & 2.47   & 2.67   & 38.47 & 61.53  & 3.00   & 0.02   & 99.89 & 0.11  \\
\hline
                      & 5  & -26.20 & -15.74 & 0.00  & 100.00 & -26.17 & -21.93 & 0.01  & 99.99 \\
10$^{11}$             & 15 & -6.91  & -3.12  & 0.02  & 99.98  & -7.05  & -7.38  & 67.83 & 32.17 \\
                      & 30 & -1.37  & -0.08  & 4.80  & 95.20  & -0.86  & -3.31  & 99.65 & 0.35 \\
\hline
\end{tabular}
\end{table*}
%%%%%%%%%%%%%%%%%%%%%%%%%%%%%%%%%%%%%%%%%%%%%%%%%%%%%%%%%%%%%%%%%%%%%%%%%%%%%%%%%%%%%%%%%%%%%%%%%%%%%%%%%%%%%%
\end{document}